\documentclass[11pt]{article}

\usepackage[preprint]{acl}

\usepackage{times}
\usepackage{latexsym}
\usepackage[T1]{fontenc}
\usepackage[utf8]{inputenc}
\usepackage{microtype}
\usepackage{inconsolata}
\usepackage{graphicx}
\usepackage{booktabs}
\usepackage{amsmath}
\usepackage{amssymb}
\usepackage{tikz}
\usepackage{tabularx}
\usepackage{listings}
\usepackage{enumitem}
\usetikzlibrary{calc,positioning}

\title{Reading Between the Citations: A Typed Claim Network for Scientific Literature}

\author{Ning Ding, \ Sergio J. Rodríguez Méndez, \and Pouya G. Omran \\
  Australian National University, Canberra ACT 2601, Australia \\
  https://comp.anu.edu.au/}

\begin{document}
\maketitle

\begin{abstract}

Knowledge graphs over corpora of inter-referencing documents - scholarly papers, legal opinions, policy briefs - encode the topology of reference but not its stance. The standard representation collapses a rich evaluative relation into an untyped edge, losing the very content that supports community-level queries about how one document is received by another. We propose the \textit{claim network}: a representational pattern in which each cross-document reference is reified as a typed claim, carrying source, target, claim text, and a four-class stance label grounded in the citation-intent literature. We give a construction pipeline applicable to any corpus of scholarly inter-referencing documents and instantiate it on a corpus of 127 papers in 3D point cloud semantic segmentation, producing a network of 8{,}260 typed claims. Three downstream task families demonstrate what the network enables: retrieval signal augmentation, aggregated-stance summarisation, and topological analytics. Head-to-head evaluation against standard Retrieval-Augmented Generation (RAG) baselines shows that the gain over flat retrieval is the gain from the right intermediate representation rather than the wrong one.

\end{abstract}


\section{Introduction}
\label{sec:introduction}

Many document corpora are held together by inter-document reference: their members cite one another and, in citing, express stances toward what they cite. Scholarly papers are the canonical case, but the same structural pattern appears in legal opinions, patents, and policy documents. Knowledge graphs organise such corpora into structured representations supporting search, recommendation, question answering, and synthesis~\citep{verma2023scholarly,abu2021domain}. The recent emergence of retrieval-augmented language models has further raised their practical value: when a generator is grounded in structured evidence rather than opaque text retrieval, both accuracy and auditability improve~\citep{lewis2020retrieval,edge2024graphrag}.

\begin{figure}[t]
\centering
\begin{tikzpicture}[
  paper/.style={draw, rounded corners=2pt, minimum width=1.3cm,
                minimum height=0.55cm, font=\small, inner sep=2pt},
  claim/.style={draw, fill=gray!10, rounded corners=2pt, align=center,
                font=\scriptsize, text width=3.2cm, inner sep=3pt},
  >=stealth, node distance=2.6cm
]
\node[paper] (a1) {Paper $A$};
\node[paper, right=of a1] (b1) {Paper $B$};
\draw[->] (a1) -- node[font=\scriptsize, above] {\textit{cites}} (b1);
\node[font=\footnotesize\itshape,
      below=0.35cm of $(a1)!0.5!(b1)$] {(a) Untyped citation edge};

\node[paper, below=2.0cm of a1] (a2) {Paper $A$};
\node[paper, right=of a2] (b2) {Paper $B$};
\node[claim, below=0.55cm of $(a2)!0.5!(b2)$] (c) {%
  \textbf{\textsc{Critique}}\\[2pt]
  ``\textit{$B$'s kernel-point design scales poorly to large scenes.}''\\[2pt]
  {\scriptsize\textit{provenance:} \S4, \P3 of $A$}};
\draw[->] (a2.south) -- (c.north west);
\draw[->] (c.north east) -- (b2.south);
\node[font=\footnotesize\itshape, below=0.3cm of c]
  {(b) Reified typed claim};
\end{tikzpicture}
\caption{From an untyped citation edge (a) to a reified typed claim (b).
The reified claim carries a type label, an extracted claim sentence, and
provenance back to the citing paragraph in paper~$A$.}
\label{fig:claim_vs_citation}
\end{figure}
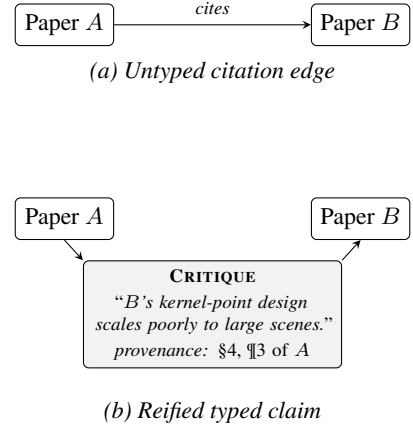

A persistent limitation of these graphs, however, is that they encode the \textit{topology} of reference but not its \textit{content}. The standard representation is an untyped edge: document A references document B. What is lost is the \textit{stance} that A takes toward B --- whether A adopts B's position, critiques its limitations, or compares its own contribution against B as a benchmark. These stances are abundant in citing prose and constitute the substrate of community evaluation, but they remain trapped in unstructured text. Downstream systems consuming the knowledge graph see only that A referenced B; the evaluative content of that reference is invisible. Figure~\ref{fig:claim_vs_citation} illustrates this contrast.

This is more than a representational inconvenience. A class of natural questions about such corpora --- \textit{``how has the community evaluated document $X$''}, \textit{``which document is most influential within a given line of work''}, \textit{``which documents bridge two areas''} --- requires aggregating evaluative stances across many references. Without typed reference, these questions are answerable only by reading every relevant document and aggregating manually. Recent work on retrieval-augmented language models partially addresses the textual side~\citep{li2025searcho1}, but flat vector retrieval over chunks cannot return objects such as \textit{``the typed set of critiques of $X$''} as a complete result, because no such object is localised in the corpus text.

We address this limitation by proposing the \textit{claim network}: a representational pattern in which each cross-document reference is reified as a typed claim carrying explicit source, target, claim text, and a type label. The construction pipeline takes any corpus of inter-referencing documents, decomposes each into a hierarchical representation with resolved references, aggregates citing context at the document-pair level, and applies a structured-output language-model step to assign each pair a typed claim drawn from a four-class taxonomy (\textsc{Critique}, \textsc{Adoption}, \textsc{Benchmark}, \textsc{Neutral}) grounded in the citation-intent literature. We instantiate the pipeline on a corpus of 127 papers in 3D point cloud semantic segmentation, using the Deep Document Model (DDM)~\citep{jia2024ddm} as the hierarchical substrate, obtaining a network of 8{,}260 typed claims. On this network we demonstrate three downstream task families that typed reference structure enables --- retrieval signal augmentation, aggregated-stance summarisation, and topological analytics --- instantiated as claim-conditioned question answering, consensus report generation, and graph analytics over the scholarly corpus.

This gap is sharpest in retrieval-augmented systems. At generation time, a language model reading a citation context can usually infer whether the cited work is being adopted, critiqued, benchmarked against, or merely mentioned. At retrieval time, by contrast --- and especially at the graph level on which retrieval is grounded --- this information is compressed into a single untyped edge. The specific context in which one paper discusses another, the textual evidence supporting that relation, and the evaluative stance expressed by the citing paper are all collapsed before retrieval even begins. The claim network closes this gap by reifying each citation as a structured object: a relation from a source paper to a target paper, anchored in a citation context, supported by an extracted claim sentence, and labelled with an explicit relation type.

\paragraph{Contributions.} (1) We formulate the \textit{claim network} as a general representational pattern for corpora of evaluative inter-referencing documents, together with a four-class stance taxonomy grounded in the citation-intent literature. (2) We give a construction pipeline applicable to any corpus of scholarly inter-referencing documents that operates over a hierarchical document decomposition with resolved references, and a Large Language Model (LLM)-based extraction stage that produces typed claims from per-pair citing contexts. (3) We instantiate the pipeline on a 127-paper corpus in 3D point cloud semantic segmentation, releasing the resulting network of 8{,}260 typed claims together with the underlying DDM corpus. (4) We demonstrate three downstream task families --- retrieval signal augmentation, aggregated-stance summarisation, and topological analytics --- and evaluate them under head-to-head LLM-as-judge comparison against retrieval-augmented baselines. Where prior work that consumes citation intent classification (CIC) labels downstream targets a single task family (\citealp{bezerra2025leveraging} for citation-network analytics under retyped edges; \citealp{ma2025citation} for citation recommendation conditioned on rhetorical zone) or operates over pre-authored typed triples rather than evaluative stances extracted from citing prose \citep{taffa2023leveraging}, our contribution is the combination of
persistent reified per-pair claims, two-axis (stance and attitude)
labeling, and evaluation across three task families on the same
artefact. (5) We audit the LLM extractor against both a held-out human
spot-check on 150 pair-windows from the corpus itself (stance Macro-F1 = $0.892$, per-label $\kappa \in [0.81, 0.90]$; attitude $\kappa = 0.735$) and the SciCite~\citep{cohan2019scicite} citation-intent benchmark under our taxonomy mapping (\textsc{Direct}-vs-\textsc{Mapped} internal $\kappa = 0.838$), separating extractor noise from the downstream effects measured in \S\ref{sec:experiments}.

\section{Related Work}
\label{sec:related_work}

The claim network sits at the intersection of three literatures, and we organise this section around each in turn. \S\ref{sec:rw_kg} surveys scholarly knowledge graphs and the hierarchical document decomposition work that supplies the substrate our construction pipeline operates over. \S\ref{sec:rw_intent} surveys citation intent classification and argument mining, the lineage from which our four-class stance taxonomy and three-valued attitude axis inherit. \S\ref{sec:rw_rag} surveys retrieval-augmented generation, against which the three downstream tasks of \S\ref{sec:tasks} are evaluated. After each subsection we make explicit what the claim network adds beyond the surveyed work.

\subsection{Scholarly Knowledge Graphs}
\label{sec:rw_kg}

Bibliographic graphs of papers, authors, and venues have a long history in scientometrics and citation recommendation~\citep{verma2023scholarly}. More recent work pushes for fine-grained representation, decomposing papers into their constituent sections, paragraphs, and sentences. The Deep Document Model and the related Document Object Model Ontology~\citep{jia2024ddm} represent academic papers as hierarchical trees of typed nodes, with explicit reference and citation entities linking sentences to invoked works. This fine-grained decomposition has been shown to improve question answering and information retrieval over scientific corpora compared to coarse-grained chunking~\citep{jia2024ddm}.

A separate line of work attempts to type the relations between scholarly entities. The Open Research Knowledge Graph (ORKG) represents research contributions as structured comparisons over typed properties~\citep{jaradeh2019orkg}, and Nanopublications~\citep{groth2010nanopublication} model individual scientific assertions as first-class graph entities with provenance. These systems share with our work the goal of moving beyond untyped citation, but they require manual or semi-automated assertion authoring and target the structured representation of \textit{scientific findings} rather than the \textit{evaluative relations between findings}. Citations themselves remain untyped in their schemas.

A third line uses knowledge-graph structure to augment retrieval-augmented language models. GraphRAG~\citep{edge2024graphrag} constructs entity-relation graphs from text corpora and uses community detection to support query-focused summarization. Other systems integrate scholarly knowledge graphs with LLMs to improve factual grounding~\citep{li2024corpuslm} or support deep research workflows~\citep{li2025webthinker}. In all of these systems, citations between papers --- where they appear at all --- are untyped edges.

Our work extends this lineage by reifying inter-document references themselves as typed claims, producing a representation that supports queries about evaluative stance rather than only reference topology or assertion content. The scholarly knowledge graph is the natural first instantiation, and the construction transfers in principle to any corpus of scholarly inter-referencing documents; cross-domain transfer (to legal opinions, patent prior-art chains, or policy documents) is consistent with the construction but is not demonstrated by the present work.

\subsection{Citation Intent Classification and Argument Mining}
\label{sec:rw_intent}

Classifying the rhetorical role of citations has been studied in two adjacent research traditions. Argumentative zoning~\citep{teufel2002summarizing,teufel2009towards} annotates the discourse function of sentences in scientific papers (\textit{Background}, \textit{Own}, \textit{Contrast}, etc.), with citations classified by their role in the surrounding argument. Citation intent classification, by contrast, focuses on the function of an individual citation: ACL-ARC~\citep{jurgens2018measuring} introduces a six-class scheme (\textit{Background}, \textit{Uses}, \textit{CompareOrContrast}, \textit{Motivation}, \textit{Extends}, \textit{Future}) over the ACL Anthology, and SciCite~\citep{cohan2019scicite} consolidates this into a three-class scheme (\textit{Background}, \textit{Method}, \textit{Result Comparison}) demonstrating that intent labels can be predicted from citing-context text. Subsequent work has refined the classification target with finer-grained labels~\citep{jantsch2025finecite} or treated it as joint structured prediction with scientific information extraction~\citep{viswanathan2021citation}. The broader scientific argument mining literature~\citep{lauscher2018argument,stab2017parsing} situates these classifications within models of argument structure across full papers. 

A more recent wave of citation-intent work extends this lineage
along several axes. \citet{lahiri2023citeprompt} reformulate CIC as
a cloze-style prompting task on pretrained language models,
eliminating the need for task-specific classification heads.
\citet{paolini2025citefusion} fuse contextual and structural signals
at the encoder level rather than at the loss. \citet{shui2024fine}
train across heterogeneous CIC datasets jointly to mitigate
label-scheme drift. Two systems in this wave explicitly connect CIC
labels to a downstream graph or retrieval consumer:
\citet{bezerra2025leveraging} use CIC labels to retype edges in
downstream citation-network analytics, and \citet{ma2025citation}
condition a citation-recommendation system on argumentative-zone
labels at retrieval time. We share the downstream-consumer move with
the latter two but differ in three ways made precise below: the
typed claim is a persistent reified object retained in the network
rather than a transient analytics input, the labelling is two-axis
rather than single-axis, and the unit of classification is the
paragraph-aggregated context for an ordered paper pair rather than
the individual citation sentence.

\paragraph{Our taxonomy.} We adopt a four-class scheme (\textsc{Critique}, \textsc{Adoption}, \textsc{Benchmark}, \textsc{Neutral}) derived from this lineage and adapted for the downstream tasks the claim network supports; the precise correspondence to ACL-ARC and SciCite labels is given in Appendix~\ref{app:taxonomy_mapping}. The four classes are intended to capture the recurrent stances expressed in evaluative inter-document reference: a citing document may critique its target, adopt it, treat it as a comparative benchmark, or refer to it neutrally as background. Two methodological features distinguish our approach. First, the classification target is not the sentence in isolation but the paragraph-level context aggregating all references from one document to another, reflecting the unit at which evaluative stance operates. Second, where prior work that does consume CIC labels downstream targets a single task family --- citation-network analytics under retyped edges \citep{bezerra2025leveraging}, citation recommendation conditioned on rhetorical zone \citep{ma2025citation}, or Knowledge Graph Question Answering (KGQA) over pre-authored typed property triples \citep{taffa2023leveraging,jaradeh2019orkg} --- we treat the typed claim as a persistent reified object in the network and evaluate the same artefact across three task families spanning retrieval, summarisation, and analytics. Orthogonal to the four-class stance taxonomy, we additionally tag each claim with a three-valued \textit{attitude} label (\textsc{Positive}, \textsc{Negative}, \textsc{Neutral}) capturing the evaluative valence of the citing prose, generalising the sentence-level citation-sentiment annotation introduced by~\citet{athar2011sentiment} to paragraph-level aggregated context. The attitude axis is independent of the type axis: a \textsc{Benchmark} claim can be \textsc{Positive} (the source paper commends the benchmark and expresses a favorable stance), \textsc{Negative} (the source paper critiques certain aspects of the benchmark and expresses a critical stance), or \textsc{Neutral} (a head-to-head report without judgement). The two-axis representation gives downstream tasks a sharper purchase on the citing prose than either axis would alone.

\subsection{Retrieval-Augmented Generation for Scientific Question Answering}
\label{sec:rw_rag}

Retrieval-augmented generation (RAG)~\citep{lewis2020retrieval} grounds language models in external knowledge and has become standard for scientific question answering~\citep{li2025searcho1}. Several lines of work augment retrieval with structure: graph-based RAG~\citep{edge2024graphrag} uses entity-relation graphs to surface evidence at the level of communities or paths; agentic RAG~\citep{yao2022react,li2025searcho1} allows language models to issue retrieval queries autonomously during reasoning; and multi-query expansion approaches~\citep{wang2023query2doc} reformulate the query into multiple variants to improve recall, though the variants are typically generated by the language model itself.

A separate strand of scholarly RAG operates directly over typed knowledge graph (KG) triples rather than text chunks. \citet{taffa2023leveraging} translate natural- language questions into SPARQL queries against the Open Research Knowledge Graph (ORKG) \citep{jaradeh2019orkg} and verbalise the returned triples as the final answer. This is structurally the closest prior analogue to Task~1, over typed property triples rather than over typed evaluative claims. We do not run it as a Task~1 baseline because the ORKG contribution-comparison schema does not cover the 3D point cloud
segmentation corpus on which our scholarly instantiation is built; importing the corpus into a KGQA-amenable schema would itself be a non-trivial contribution. The relevant distinction with respect to our work is that ORKG triples encode pre-authored scientific findings, whereas the claim network encodes paragraph-level evaluative stances extracted at scale from citing prose.

Our Task~1 system relates closely to multi-query expansion, but its queries are drawn from a structured artifact --- the claim network --- rather than generated by the language model. This distinction matters because claim-derived queries encode evaluative phrasings that LLM-generated expansions are unlikely to produce, and because the claim network is corpus-specific in a way that LLM expansion is not. Our Task~2 system departs more sharply: it accesses the claim network directly without chunk retrieval, using typed claim lists as primary content. This is closer in spirit to graph-based RAG, but the structured object consumed is a typed set of evaluative stances rather than a community of related entities. We report a direct head-to-head against GraphRAG~\citep{edge2024graphrag} on the consensus-report task in \S\ref{sec:task2_graphrag}, together with a cost-of-inference comparison that motivates the framing.

\section{Claim Network Construction}
\label{sec:claim_network}

Given a corpus of interlinked documents whose members reference and discuss one another, our pipeline enriches the underlying reference-level knowledge graph with typed, evidence-grounded claims. The pipeline has four stages: (i) it identifies reference events between document pairs; (ii) it localises the textual context in which one document refers to another; (iii) it extracts the claim expressed in that context; and (iv) it classifies the resulting relation under an evaluative-stance taxonomy. The output --- a claim network --- can be consumed by downstream systems for question answering, corpus-level consensus generation, graph analytics, and visualisation. The construction is described in scholarly-neutral terms --- an inter-referencing corpus, a hierarchical decomposition, per-pair aggregation, structured extraction --- so that the relevant primitives are visible at construction time; we instantiate and evaluate it on a scholarly corpus and treat cross-domain instantiation as future work.

\begin{figure}[htbp]
    \centering                            
    \includegraphics[width=0.5\textwidth]{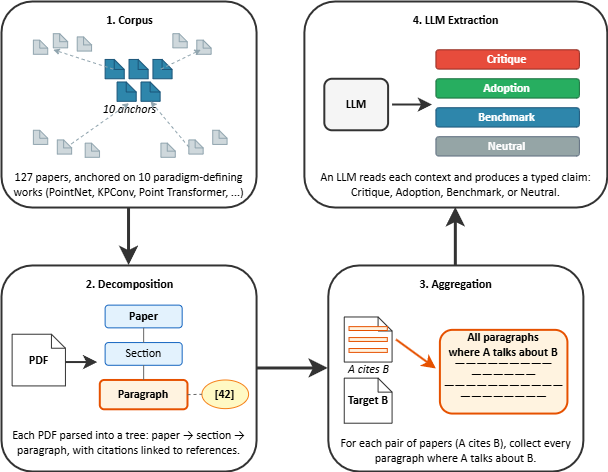}
    \caption{The four-stage construction pipeline, instantiated on a scholarly corpus. (1) A 127-paper corpus is assembled by star-shaped expansion around ten paradigm-defining anchors. (2) Each PDF is decomposed into a hierarchical document representation with resolved references. (3) For each ordered pair of papers, every paragraph in which the source discusses the target is aggregated into a single context. (4) An LLM reads each context and produces a typed claim under the four-class taxonomy.}
    \label{fig:pipeline}
\end{figure}

We instantiate the pipeline on a scholarly corpus, using the Deep Document Model~\citep{jia2024ddm} as the hierarchical substrate against which citation contexts are localised. Sections~\ref{sec:corpus} describe the four stages --- corpus selection, document decomposition, claim extraction, and quality validation --- as instantiated for a 127-paper corpus in 3D point cloud semantic segmentation. Each extracted claim carries two independent labels: a multi-label stance set drawn from a four-class taxonomy (\textsc{Critique}, \textsc{Adoption}, \textsc{Benchmark}, \textsc{Neutral}) capturing what the source paper \emph{does} with its target, and a single-label attitude (\textsc{Positive}, \textsc{Negative}, \textsc{Neutral}) capturing how the source paper \emph{feels} about its target. The resulting network contains 8{,}260 typed claims; statistics are reported in \S\ref{sec:stats}. Figure~\ref{fig:pipeline} summarises the four stages.

\subsection{Corpus}
\label{sec:corpus}

The corpus centers on 3D point cloud semantic segmentation, a domain that exhibits paradigmatic diversity (point-based, voxel-based, graph-based, transformer-based, self-supervised) while remaining tightly coupled through shared citations. We construct the corpus through star-shaped expansion around ten manually selected core papers chosen to span the major architectural paradigms: PointNet, PointNet++, DGCNN, KPConv, RandLA-Net, MinkowskiNet, Point Transformer, Point-MAE, PointNeXt, and Uni3D. These ten anchors form a paradigm-balanced cross-section rather than a top-cited list.

For each core paper, we use the Semantic Scholar API to retrieve its eight most-cited \textit{inbound citations} (papers citing the core, ranked by their own citation counts) and eight most-cited \textit{outbound references} (papers the core cites, similarly ranked), restricting to those with arXiv identifiers so that full PDFs are available. After deduplication, the corpus contains 127 unique papers. The star-shaped construction produces a corpus densely connected around its anchors rather than uniformly sampled across the field; quantitative findings derived from the network describe the local citation neighborhood of the anchors.

\subsection{Document Decomposition}
\label{sec:ddm}

As the hierarchical document decomposition required by the pipeline, we adopt the Deep Document Model (DDM)~\citep{jia2024ddm}, which is particularly well-matched to scholarly papers because it natively represents \textit{Reference} entities and \textit{Citation} edges. Each PDF in the corpus is converted to a DDM representation: a hierarchical tree of typed nodes (\textit{Paper} $\rightarrow$ \textit{Section} $\rightarrow$ \textit{Paragraph} $\rightarrow$ \textit{Sentence}) with \textit{Reference} entities and \textit{Citation} edges linking sentences to the references they invoke. Conversion proceeds in two stages. The PDF is first rendered to Markdown using \texttt{marker}\footnote{\url{https://github.com/datalab-to/marker}}, an open-source layout-aware PDF-to-Markdown converter. A DDM-specific parser then walks the Markdown output to recover the section hierarchy from heading levels, identify paragraph and sentence boundaries, and extract the bibliography as a list of \textit{Reference} entities; inline citation markers in body paragraphs are resolved against this reference list to produce the \textit{Citation} edges. The complete representation is serialised to Resource Description Framework (RDF) and stored in a GraphDB triplestore. Paragraph-level localisation of citations is the property of this substrate on which the rest of the pipeline depends: it is what allows downstream extraction to distinguish a critique from an adoption.

\subsection{Extraction Pipeline}
\label{sec:extraction}

Claims are extracted along two independent axes introduced in \S\ref{sec:rw_intent}: the four-class stance taxonomy (\textsc{Critique}, \textsc{Adoption}, \textsc{Benchmark}, \textsc{Neutral}), which the extractor treats as a multi-label set so that a single claim can simultaneously be both an \textsc{Adoption} and a \textsc{Benchmark}; and the three-valued attitude label (\textsc{Positive}, \textsc{Negative}, \textsc{Neutral}), which the extractor treats as a single-label classification. The two axes are predicted jointly in one structured-output call rather than sequentially, so the model sees the citing prose under one decoding pass and emits both labels for each window. The pipeline has three stages.

\paragraph{Context aggregation.} For each ordered pair $(p_s, p_t)$ in which the source paper $p_s$ cites the target paper $p_t$ at least once, we issue a SPARQL Protocol and RDF Query Language (SPARQL) query against GraphDB that traverses \textit{Paper} $\rightarrow$ \textit{Section} $\rightarrow$ \textit{Paragraph} $\rightarrow$ \textit{Citation} chains and returns every paragraph in $p_s$ that cites $p_t$. Bibliography sections are excluded. Each retrieved paragraph forms a citation window, and the pair is associated with the list of all such windows. Aggregating at the pair level reflects the unit at which evaluative stance operates: a paper takes a coherent stance toward another paper, even when articulated across multiple paragraphs.

\paragraph{Noise filtering.} Paragraphs returned by the SPARQL traversal occasionally contain conversion artifacts: page-anchor HTML spans that \texttt{marker} embeds inline in the Markdown output, surviving bibliography fragments that escaped the heading-level section filter, and degenerate strings under twenty characters. We apply a regular-expression filter discarding these patterns. Pairs whose windows are entirely filtered out are skipped.

\paragraph{LLM extraction with structured output.} For each surviving pair, we issue a single prompt to Qwen3-Max via the DashScope API, providing the source paper title, target reference text, and concatenated windows. The model produces, for each window, a specific claim sentence, a multi-label set of stance labels drawn from the four-class taxonomy, and a single-label attitude. The schema enforces the cardinality difference between the two axes: the stance set is required to be non-empty and may carry up to four labels with \textsc{Neutral} mutually exclusive of the other three (a prompt-level constraint rather than a schema-level one), while the attitude is required to be exactly one of the three valid values. Outputs are constrained to JSON via the model's structured-output mode and validated against a Pydantic schema. The full prompt, schema, and implementation details are given in Appendix~\ref{app:implementation}.

\paragraph{Reproducibility configuration.} Table~\ref{tab:repro} records
the decoding parameters and operational settings used at extraction time.
Together with the prompt and Pydantic schema reproduced in Appendix~\ref{app:implementation}, these constitute the reproducibility
floor for the construction pipeline.

\begin{table}[t]
\centering
\small
\setlength{\tabcolsep}{3.5pt}
\begin{tabularx}{\columnwidth}{@{} l X @{}}
\toprule
\textbf{Setting} & \textbf{Value} \\
\midrule
Model & Qwen3-Max \\
\addlinespace
Endpoint & DashScope OpenAI-compatible \texttt{/compatible-mode/v1} \\
\addlinespace
Temperature & 0.0 (greedy decoding) \\
\addlinespace
\texttt{top\_p} & DashScope default (not overridden) \\
\addlinespace
\texttt{max\_tokens} & DashScope default (not overridden) \\
\addlinespace
\texttt{response\_format} & \texttt{\{"type": "json\_object"\}} \\
\addlinespace
Retries on schema / API failure & 3 attempts; flat 2\,s sleep between \\
\addlinespace
Quota / billing errors & raised immediately (no retry) \\
\addlinespace
Seed strategy & none; determinism via \texttt{temperature=0.0} only \\
\addlinespace
Concurrency cap & \texttt{asyncio.Semaphore(5)} \\
\addlinespace
Prompt / schema version & released with code (App.~\ref{app:b1_prompt}, \ref{app:b2_schema}) \\
\bottomrule
\end{tabularx}
\caption{Reproducibility configuration for the extraction pipeline of
\S\ref{sec:extraction}. The Qwen3-Max identifier is the DashScope alias
rather than a dated snapshot; we record this here so that future re-runs
against a refreshed alias are recognised as a reproducibility hazard
rather than a silent drift.}
\label{tab:repro}
\end{table}



\subsection{Resulting Network}
\label{sec:stats}

The pipeline produces a claim network of 8{,}260 typed claims over 127
papers, derived from a DDM graph containing 3{,}662 sections, 19{,}358
paragraphs, and 4{,}210 reference entities (Table~\ref{tab:stats}).

\begin{table}[t]
\centering
\small
\begin{tabular}{lr}
\toprule
\textbf{Statistic} & \textbf{Value} \\
\midrule
Papers / Sections / Paragraphs & 127 / 3{,}662 / 19{,}358 \\
Reference entities & 4{,}210 \\
\midrule
Total claims & 8{,}260 \\
\midrule
\multicolumn{2}{l}{\textit{Stance labels (multi-label, sum $>$ total)}} \\
\quad Critique & 723 \\
\quad Adoption & 1474 \\
\quad Benchmark & 2{,}489 \\
\quad Neutral & 4{,}206 \\
\midrule
\multicolumn{2}{l}{\textit{Attitude labels (single-label, sum $=$ total)}} \\
\quad Positive & 2286 \\
\quad Negative & 786 \\
\quad Neutral & 5188 \\
\midrule
Median in-claims per paper & 8 \\
Median out-claims per paper & 46 \\
\bottomrule
\end{tabular}
\caption{DDM corpus and claim network statistics. The stance axis is
multi-label; the attitude axis is single-label.}
\label{tab:stats}
\end{table}

\textit{Neutral} claims constitute roughly 53\% of the network, consistent with the well-documented prevalence of non-evaluative citation~\citep{cohan2019scicite}. Among evaluative classes, \textit{Benchmark} dominates at 31\%, reflecting a community in which empirical comparison against established baselines is the dominant mode of citation. \textit{Critique} is the rarest class at 7\%. The most-cited target is \textit{4D Spatio-Temporal ConvNets: Minkowski Convolutional Neural Networks}, structurally reflecting its role as the canonical voxel-based baseline. The most prolific source is the survey \textit{Aligning Cyber Space with Physical World: A Comprehensive Survey on Embodied AI}. The attitude distribution is informative as a cross-check on the stance distribution. \textsc{Negative} attitude is more prevalent than \textsc{Critique} stance, because a claim can carry a \textsc{Negative} attitude under a non-\textsc{Critique} stance - most commonly a \textsc{Benchmark} claim in which the source paper reports outperforming the target. \textsc{Positive} attitude similarly extends beyond pure \textsc{Adoption}: a \textsc{Benchmark} claim that places the source paper favourably against the target carries \textsc{Positive} attitude under a non-\textsc{Adoption} stance. The two axes therefore capture overlapping but non-redundant signal, which is the property that makes them independently useful in the downstream tasks of \S\ref{sec:tasks}. The accuracy of the extractor that produced these labels is reported separately in \S\ref{sec:extractor_validation}.

\subsection{Extractor Validation}
\label{sec:extractor_validation}

The downstream analyses in \S\ref{sec:experiments} consume the typed
claim network as if its labels were ground truth. A reader cannot
separate the downstream effects from extractor noise without a direct
measurement of how accurate those labels are. We report two
complementary evaluations: an in-corpus human spot-check on a sample
drawn from the network itself, and a cross-scheme transferability
check against SciCite~\citep{cohan2019scicite} under the
Appendix~\ref{app:taxonomy_mapping} mapping.

\paragraph{In-corpus human spot-check.}
We sampled 150 pair-windows uniformly from the post-filter window set
of \S\ref{sec:extraction} and hand-labelled each by the first author
under the same definitions and prompt instructions given to the
extractor. Annotation was single-annotator and is therefore an
agreement-with-the-extractor measurement rather than an inter-annotator-
agreement measurement; we return to this limitation at the end of the
subsection. Table~\ref{tab:extractor_audit_stance} reports per-class
precision, recall, F1, and Cohen's $\kappa$ on the stance axis
(multi-label, evaluated per label as a binary classification);
Table~\ref{tab:extractor_audit_attitude} reports the 3$\times$3
confusion matrix on the single-label attitude axis with overall
accuracy, macro-F1, and $\kappa$.

\begin{table}[t]
\centering
\small
\setlength{\tabcolsep}{4pt}
\begin{tabularx}{\columnwidth}{@{} l c c c c @{}}
\toprule
\textbf{Stance label} & \textbf{P} & \textbf{R} & \textbf{F1} & \textbf{$\kappa$} \\
\midrule
Adoption  & 0.775 & 1.000 & 0.873 & 0.835 \\
Benchmark & 0.875 & 0.921 & 0.897 & 0.861 \\
Critique  & 0.891 & 0.976 & 0.932 & 0.904 \\
Neutral   & 0.975 & 0.780 & 0.867 & 0.811 \\
\midrule
\textit{Macro-F1} & \multicolumn{4}{c}{\textit{0.892}} \\
\textit{Exact set-match accuracy} & \multicolumn{4}{c}{\textit{0.853}} \\
\bottomrule
\end{tabularx}
\caption{Extractor versus human stance labels on 150 hand-annotated
pair-windows. Stance is multi-label and each label is evaluated as a
binary classification (presence vs absence). Per-label $\kappa$ is
in the ``almost perfect'' range
($0.81$--$1.00$;~\citealp{landis1977measurement}) for all four
classes; exact set-match agreement is 85.3\%.}
\label{tab:extractor_audit_stance}
\end{table}

\begin{table}[t]
\centering
\small
\setlength{\tabcolsep}{8pt}
\begin{tabularx}{\columnwidth}{@{} l c c c c @{}}
\toprule
\textbf{Human \textbackslash{} LLM} & \textbf{Pos} & \textbf{Neg} & \textbf{Neu} & \textbf{Total} \\
\midrule
Positive & 30 & 2 & 3 & 35 \\
Negative & 2 & 39 & 0 & 41 \\
Neutral  & 16 & 3 & 55 & 74 \\
\midrule
Total    & 48 & 44 & 58 & 150 \\
\bottomrule
\end{tabularx}
\caption{Extractor versus human attitude labels on the same 150
windows. Overall accuracy $=0.827$, macro-F1 $=0.825$, Cohen's
$\kappa = 0.735$ (substantial agreement per
\citealp{landis1977measurement}). Per-class F1: Positive $0.723$,
Negative $0.918$, Neutral $0.833$.}
\label{tab:extractor_audit_attitude}
\end{table}

The stance axis is in ``almost perfect'' agreement on every class
under the Landis--Koch interpretation: per-label $\kappa$ ranges from
$0.811$ (Neutral) to $0.904$ (Critique), and Macro-F1 is $0.892$. The
attitude axis is in ``substantial'' agreement overall ($\kappa
= 0.735$) with Negative the strongest class ($\kappa$-equivalent F1
$0.918$) and Positive the weakest ($0.723$). The dominant disagreement
pattern on both axes is asymmetric in the same direction: the
extractor is slightly more willing than the human annotator to call a
citation evaluative. On stance, this surfaces as the LLM under-
predicting Neutral (recall $0.780$, with 11 of the 50 human-Neutral
windows assigned an evaluative label by the LLM); on attitude, as
the LLM over-predicting Positive (precision $0.625$, with 16 of the
74 human-Neutral windows assigned Positive by the LLM). The two
patterns are not independent: a window the LLM reads as a soft
adoption signal will tend to receive an Adoption stance and a
Positive attitude jointly. The bias is mild and consistent. It does
not invalidate the downstream tasks of \S\ref{sec:experiments} but
should be read forward into them: the network slightly
over-represents evaluative content relative to a stricter human
gold, which works in favour of the \S\ref{sec:experiment_task2}
\textit{Information Density} result rather than against it.

\paragraph{Cross-scheme transferability against SciCite.}
We additionally evaluated the extractor against the SciCite test set
(1{,}859 sentences) under the Appendix~\ref{app:taxonomy_mapping}
mapping. Two prediction modes were compared. \textsc{Direct}: the
extractor is prompted to assign each sentence one of SciCite's three
labels (\textit{background}, \textit{method}, \textit{result})
directly. \textsc{Mapped}: the extractor produces a four-class
\texttt{claim\_types} label under its usual prompt, and the label is
mapped to a SciCite class via the correspondence in
Table~\ref{tab:taxonomy_mapping}. Results are given in
Table~\ref{tab:scicite_eval}.

\begin{table}[t]
\centering
\small
\setlength{\tabcolsep}{4pt}
\begin{tabularx}{\columnwidth}{@{} l c c c @{}}
\toprule
\textbf{Comparison} & \textbf{Acc} & \textbf{Macro-F1} & \textbf{$\kappa$} \\
\midrule
\textsc{Direct} vs SciCite gold       & 0.706 & 0.661 & 0.505 \\
\textsc{Mapped} vs SciCite gold       & 0.702 & 0.644 & 0.480 \\
\midrule
\textsc{Direct} vs \textsc{Mapped}    & 0.906 & --- & 0.838 \\
\bottomrule
\end{tabularx}
\caption{Extractor against SciCite gold under the
Appendix~\ref{app:taxonomy_mapping} mapping. The two prediction modes
reach essentially identical agreement with gold; their internal
agreement is in the ``almost perfect'' range. The shared gap from
gold is therefore not attributable to the taxonomy mapping but to
the underlying task: SciCite operates at the sentence level on a
heterogeneous biomedical/NLP corpus, while our extractor was
designed for paragraph-aggregated contexts on a single scientific
subfield.}
\label{tab:scicite_eval}
\end{table}

Two observations follow. First, the taxonomy mapping is faithful as
a conceptual translation: \textsc{Direct} and \textsc{Mapped} predictions agree with each other at $\kappa = 0.838$. Whatever the extractor decides to call a claim under its native four-class taxonomy maps consistently onto the same SciCite label that it would assign if asked directly. Second, both modes reach the same moderate agreement with SciCite gold ($\kappa \approx 0.49$). The gap from gold is therefore not introduced by the mapping; it is shared between the two modes and reflects the unit mismatch (SciCite is sentence-level annotation; our extractor was tuned for paragraph-aggregated context) and the domain mismatch (SciCite spans biomedical and natural language processing (NLP) papers; our corpus is 3D point cloud semantic segmentation). The intent of this evaluation is therefore not to claim parity with SciCite-tuned classifiers but to validate that the Appendix~\ref{app:taxonomy_mapping} mapping is faithful and that the extractor's predictions, projected onto SciCite's label space, behave coherently rather than erratically.

\paragraph{Scope and limitations of the audit.}
Three limitations of the present audit should be acknowledged. First,
the human annotation is single-annotator. The numbers above are
agreement between the extractor and one trained annotator, not an
inter-annotator agreement among multiple independent annotators; the
$\kappa$ values cannot be decomposed into ``extractor noise'' versus
``annotation noise.'' Second, both the human annotator and the
extractor saw the same prompt and the same label definitions, so the
audit measures execution under shared instructions rather than
agreement between independent operationalisations of the taxonomy.
Third, the SciCite transferability evaluation is informative about
the mapping and about the unit-level behaviour of the extractor on
sentence-scale inputs, but does not directly evaluate the network's
paragraph-aggregated claims, which are the unit on which the
downstream tasks of \S\ref{sec:experiments} operate. A two-annotator
$\kappa$ on the same 150 windows, and a paragraph-aggregated
re-annotation of a held-out portion of the corpus, are the two
natural follow-ups; both are noted in \S\ref{sec:limitations}.

\section{Downstream Tasks}
\label{sec:tasks}

Typed reference structure unlocks three families of downstream task that neither untyped citation graphs nor flat vector retrieval can support. \textit{Retrieval signal augmentation} uses typed claims as auxiliary retrieval queries, surfacing evidence that lexically matches evaluative phrasing rather than only the original information need. \textit{Aggregated-stance summarisation} consumes the set of incoming typed claims for a target document as the primary input to a structured generation prompt, returning an object that no single source chunk contains. \textit{Topological analytics} applies standard graph algorithms to type-restricted projections of the network, returning rankings and structural properties that are not localised in any text region. None of these capabilities is itself novel; what is novel is that a typed claim network puts them within reach for any corpus of evaluative inter-referencing documents.
We instantiate each family on the scholarly claim network: claim-conditioned question answering (\S\ref{sec:task1}), consensus report generation (\S\ref{sec:task2}), and graph analytics over the citation pattern (\S\ref{sec:task3}). All three instantiations share the same network and corpus and use the same generator (Deepseek-V3).

\subsection{Task 1: Question Answering}
\label{sec:task1}

This task instantiates retrieval signal augmentation: typed claims enter the system as auxiliary retrieval queries that shape which chunks reach the generator, but the claim text itself never enters the generation prompt. \S\ref{sec:task1} (Setup and Pipeline) gives
the multi-query retrieval pipeline and its hyperparameters; \S\ref{sec:task1_rationale} documents the two earlier strategies that
failed before we arrived at this pipeline and the design rationale
connecting them; the closing Evaluation subsubsection specifies the
question set and the head-to-head protocol.

\subsubsection{Setup and Pipeline}

This task instantiates retrieval signal augmentation: typed claims serve as auxiliary queries against the underlying chunk index, while the claim text itself never enters the generation prompt. Given a question $q$ and a chunk index $\mathcal{V}$ over the corpus, the standard retrieval-augmented baseline retrieves the top-$K$ chunks by cosine similarity to $q$ and passes them to a frozen generator $\mathcal{M}$. Our system additionally exploits a pre-computed embedding index $\mathcal{I}_{\mathcal{C}}$ over the claim sentences in the network.

The pipeline has five stages followed by a post-fusion filter. (i) The question is embedded and retrieved against $\mathcal{V}$ for the top-$N_q = 15$ chunks. (ii) The same embedding is retrieved against $\mathcal{I}_{\mathcal{C}}$ for the top-$K_c = 5$ question-relevant claims; claims whose similarity to $q$ falls below a relevance floor $\tau_{\text{floor}} = 0.25$ are dropped before retrieval, so that off-topic claims cannot inject noise into the chunk pool. (iii) Each surviving claim is treated as an additional retrieval query against $\mathcal{V}$, returning $M = 3$ chunks per claim. (iv) For each high-confidence claim --- one whose similarity to $q$ exceeds $\tau_{\text{boost}} = 0.55$ --- the question (not the claim text) is additionally retrieved against $\mathcal{V}$ restricted to the two papers the claim connects, returning a further $M = 3$ chunks per claim. This linked-paper boost exploits the bibliographic anchoring of a strong claim: when a claim reliably identifies a relation between two specific papers, the question is more usefully retrieved within that pair than across the whole corpus. (v) The question-retrieval list, the $K_c$ claim-retrieval lists, and the linked-paper lists are merged by reciprocal rank fusion (RRF) with $k = 60$, weights $w_{\text{question}} = 2.0$, $w_{\text{claim}} = 0.4$, and $w_{\text{linked}} = 0.8$. The question stream dominates the fusion deliberately, so that claim-driven retrieval contributes additional evidence rather than displacing the original information need; we discuss the calibration of these weights in \S\ref{sec:task1_rationale}.

A post-fusion anchor filter then enforces precision: a fused chunk survives only if it was retrieved by the question stream in stage (i), by a sufficiently strong claim (similarity above $\tau_{\text{anchor}} = 0.30$) in stage (iii), or by the linked-paper boost in stage (iv). Chunks introduced exclusively by weak or tail claims are discarded, which trades recall for precision in the direction \S\ref{sec:task1_rationale} motivates. The top-$K = 10$ surviving chunks are passed to $\mathcal{M}$. Claim text never enters the generation prompt; the network influences only which chunks reach the generator, and the generator's prompt and decoding configuration are identical to the baseline's.

\subsubsection{Design Rationale}
\label{sec:task1_rationale}

We arrived at multi-query retrieval after two earlier strategies failed to produce reliable improvements. Documenting them is informative about how claim networks can and cannot help retrieval.

\textbf{Claims as appended context.} Retrieving the top-$K_c$ question-relevant claims and appending their text to the generation prompt produced rubric scores statistically indistinguishable from the baseline.\footnote{Under the head-to-head protocol of \S\ref{sec:eval_protocol}, ties dominated on every rubric dimension and the judge reported justifications consistent with the generator treating injected claim text as background rather than as evidence to integrate.} Adding claims to the prompt did not change the chunks driving the answer.

\textbf{Claims as paper-level filter.} Restricting the chunk pool to chunks from papers appearing in the retrieved claims left the chunks reaching the generator identical to the baseline's.\footnote{Diagnostic measurement on the 15 evaluation questions: the top-15 question-retrieved chunks already drew from papers heavily covered by the claim network, so the endorsed-paper set and the broad-retrieval paper set overlapped almost completely.} The intervention was structurally the wrong shape: filtering operates on \textit{which papers}, but vector retrieval at this corpus size is already saturated at the paper level.

\textbf{Multi-query retrieval.} The intervention that does change the chunk pool operates on \textit{which chunks within papers}. Treating each relevant claim as a retrieval query brings evaluative phrasings into the retrieval signal directly: a claim such as ``KPConv's kernel-point parametrization is computationally expensive at scale'' surfaces chunks the original question would rank lower. This design changes which chunks reach the generator while keeping the prompt and generator identical to the baseline.

\subsubsection{Evaluation}

We evaluate on 15 hand-curated questions covering factual,
methodological, comparative, and evaluative types. Each question is
answered once by the baseline and once by our system, and the pair is
judged under the four-trial counterbalanced LLM-as-judge protocol of
\S\ref{sec:eval_protocol}. Reported win rates and p-values are pooled
across the three independent runs of the procedure described there,
under Holm--Bonferroni correction across the four rubric dimensions.

\subsection{Task 2: Consensus Report Generation}
\label{sec:task2}

This task instantiates aggregated-stance summarisation: the set of
incoming typed claims for a target paper is consumed directly as
input to the generation prompt, with no chunk retrieval. The
subsubsection that follows states the task; \textit{Systems and
Asymmetry} contrasts the chunk-retrieval baseline with the
claim-driven system and motivates the asymmetric input design that
is intrinsic to the contribution; the closing \textit{Evaluation}
subsubsection specifies the target set, rubric, and head-to-head
protocol.

\subsubsection{Setup}

This task instantiates aggregated-stance summarisation: the set of incoming typed claims for a target paper is consumed directly as input to the generation prompt, with no chunk retrieval. Given a target paper $p_t$, the goal is to produce a structured report --- \textit{Strengths}, \textit{Limitations}, \textit{Use as Baseline}, \textit{Overall Reception} --- summarising how the community has evaluated $p_t$. The answer is not factual content from any single paper but an aggregation of evaluative stances across many papers' citing prose. The unit being asked about is precisely what the claim network reifies; aggregated-stance summarisation is its natural harvesting strategy.

\subsubsection{Systems and Asymmetry}

\textbf{Baseline.} Standard RAG: a templated query is issued against the chunk index $\mathcal{V}$, the top-$K = 15$ chunks are retrieved, and $\mathcal{M}$ generates the report.

\textbf{Ours.} A SPARQL query against the claim network returns every claim with target $p_t$, with \textsc{Neutral} claims excluded at retrieval time. Each remaining claim is placed under every type bucket carried in its multi-label \texttt{claim\_types} set, so that a claim labelled simultaneously as \textsc{Adoption} and \textsc{Benchmark} appears under both type headers in the generation prompt. This is the intended behaviour: a claim that is genuinely both an adoption and a benchmark should be visible to the generator under both rubrics rather than forced into one. Per-type lists are capped at $M = 15$ via Maximal Marginal Relevance (MMR) selection~\citep{carbonell1998use} configured for pure diversity (minimising the maximum cosine similarity of each new selection to the already-selected set) on the pre-computed claim embedding index of \S\ref{sec:extraction}; the cap is a working default, and an ablation on $M \in \{5, 15, 30, \text{all}\}$ showing that output quality is insensitive to the cap within this range is reported in \S\ref{sec:task2_mmr_ablation}. The grouped lists are passed to $\mathcal{M}$ under explicit type headers; no chunk retrieval occurs and the generator is identical to the baseline's. The generator is given a 500-word budget for the report.

The two systems have different inputs by design. The set of community evaluations of $p_t$ is distributed across the citing prose of every paper that has evaluated $p_t$, and identifying it requires citation-target resolution and stance classification --- exactly what the claim network provides at extraction time. Holding inputs identical would deny our system access to the structured artifact whose construction is the contribution. We hold the generator and output structure constant; the input difference is the demonstration.

\subsubsection{Evaluation}

We evaluate on 10 target papers spanning paradigms and reception profiles. Both systems produce one report per target. The rubric covers \textit{Perspective Coverage}, \textit{Balance and Fairness}, \textit{Source Diversity}, \textit{Information Density}, and \textit{Structural Coherence}. Pairwise head-to-head uses the same counterbalancing protocol as Task~1.

\subsection{Task 3: Graph Analytics}
\label{sec:task3}

This task instantiates topological analytics over the claim network:
standard graph algorithms applied to type-restricted projections of
the network return rankings whose evidence is the topology of the
citation pattern itself rather than any text region of any paper.
\textit{Capability Demonstration} explains why this task is reported
as a capability demonstration rather than a head-to-head evaluation;
\textit{Paper-Level Graph Projection} defines the multigraph the
queries operate over; \textit{Query Families} enumerates the query
types and the graph algorithm each reduces to on a type-restricted
subgraph; and \textit{Pipeline} specifies the implementation.

\subsubsection{Capability Demonstration}

This task instantiates topological analytics: standard graph algorithms operate on type-restricted projections of the claim network, and the output is a ranked list of papers under a structural criterion with the topology of the network itself as evidence. Such queries ("most influential paper in the Transformer literature", "most contested reception") have answers that are not localised in any text region of any paper; they are properties of the citation pattern across the corpus. We do not benchmark against a baseline because vector retrieval cannot produce a topological ranking, and a parametric language model can produce one only by drawing on training-data priors unrelated to our specific 127-paper corpus.

\subsubsection{Paper-Level Graph Projection}

The claim network is projected into a paper-level directed multigraph $G = (V, E)$. Each paper becomes a node. For each ordered pair $(p_s, p_t)$ with at least one claim, we create an edge carrying typed counts $n^\tau_{(s,t)}$ for $\tau \in \{\textsc{Cr}, \textsc{Ad}, \textsc{Be}, \textsc{Ne}\}$ and the total $n_{(s,t)}$. Topic-scoped queries restrict $G$ to a hand-curated paper subset $\mathcal{T}(\text{topic}) \subseteq \mathcal{P}$ before any algorithm is applied. We adopt hand-curated topic labels because the corpus is small enough for manual labeling to be reliable, and because closed topic vocabularies make scoped queries unambiguous.

\subsubsection{Query Families}

Six query families map to standard graph operations over typed-edge projections of the paper-level graph. The first four return ranked lists of papers under different structural criteria; the fifth returns paired rankings under a polarity score; the sixth returns the typed citation behaviour of an entire sub-community.

\textbf{1. Influence.} Weighted PageRank computed over the
\textsc{Adoption} subgraph of the paper graph, with edges weighted by
adoption-claim count $n^{\textsc{Ad}}_{(s,t)}$, then filtered to the
topic subset $\mathcal{T}(\text{topic})$ before reporting the top-$k$.
PageRank is preferred over raw weighted in-degree because it rewards
being adopted by influential papers, mitigating the artefact whereby
survey papers dominate raw-count rankings; restricting to the \textsc{Adoption} subgraph is what isolates the influence-by-adoption
signal from undifferentiated reference traffic, and is the move the
typed claim network makes possible.

\textbf{2. Most Critiqued.} Sum of $n^{\textsc{Cr}}_{(s,t)}$ over incoming edges at each node $t$, filtered to the topic subset, top-$k$ by score. This is the natural read of the \textsc{Critique} subgraph in-degree.

\textbf{3. Benchmark Status.} Sum of $n^{\textsc{Be}}_{(s,t)}$ over incoming edges at each node $t$, otherwise as above. Distinct from \textit{Influence}: a paper can carry high benchmark in-degree without high adoption (the paradigmatic case is a strong but superseded baseline) or vice versa.

\textbf{4. Most Endorsed / Most Rejected.} Sums of incoming attitude-weighted counts $n^{+}_{(s,t)}$ and $n^{-}_{(s,t)}$ respectively. The attitude axis is independent of the claim-type axis (\S\ref{sec:extraction}), so these rankings are not redundant with families 2 and 3: a paper can be widely benchmarked without being widely endorsed (mixed-result benchmarking), and a paper can be widely endorsed without being widely benchmarked (foundational adoption without head-to-head comparison).

\textbf{5. Polarity (Loved vs Controversial).} For each node $t$ with at least $\theta = 5$ incoming claims, compute the polarity score:
$$\rho(t) = \frac{n^{\textsc{Ad}}(t) - n^{\textsc{Cr}}(t)}{n_{\text{total}}(t)} \in [-1, 1].$$
Top-$k$ at the positive extreme returns the most \textit{loved} papers (broadly adopted with little critique); top-$k$ at the negative extreme returns the most \textit{controversial} (central but contested). The threshold $\theta$ excludes sparse targets whose polarity score is dominated by a handful of claims.

\textbf{6. Topic Citation Behaviour.} For each topic $\mathcal{T}$, sum $n^{\tau}_{(s,t)}$ for $\tau \in \{\textsc{Cr}, \textsc{Ad}, \textsc{Be}, \textsc{Ne}\}$ over edges \emph{outgoing} from nodes in $\mathcal{T}$, and report the percentage split across the four types together with the paper count and total outgoing claim count for the topic. This is the only query family in the set that operates on edge \emph{distribution} rather than on a per-node ranking, and the only one whose output is a property of a sub-community rather than of an individual paper. It answers a question the other families cannot: \emph{does this sub-community cite differently from that one?} - e.g. is the transformer-based sub-community more benchmark-driven than the point-based one, or more critique-driven than the graph-based one.

\textbf{Bridge papers.} A separate query family targets boundary nodes that connect two topic subsets. For each ordered pair $(\mathcal{T}_1, \mathcal{T}_2)$ in a fixed set of paradigm pairs (Transformer / Point-based, Transformer / CNN, Graph / Point-based, Self-supervised / Transformer), each node's bridge score is the minimum of its neighbour count in $\mathcal{T}_1$ and its neighbour count in $\mathcal{T}_2$, computed over the union of in- and out-neighbours. Ties are broken by the sum of the two counts. The use of $\min(a, b)$ rather than $a + b$ or betweenness centrality is deliberate: it forces actually two-sided coverage, so that a node connected heavily to one topic and weakly to the other does not outrank a node connected moderately to both. Betweenness would happily reward one-sided shortcut nodes; $\min(a, b)$ does not.

The contribution is not in the algorithms themselves but in the demonstration that, once the typed claim network exists, an entire class of corpus-level questions reduces to applying these standard operations with type-restricted edge weights.

\subsubsection{Pipeline}

A natural-language query is parsed into a structured specification (\textit{query family}, \textit{topic scope}, \textit{depth} $k$) by a single LLM call with schema-validated output. The corresponding subgraph or projection is computed, the algorithm is run, and the ranked list is returned. For influence queries, the top-ranked paper is additionally annotated with a \textit{reception profile}: distinct citing-paper count, per-type and per-attitude incoming counts, the polarity score, and a single representative critique drawn from the underlying claim network (the longest \textsc{Critique} claim with the ranked paper as target, truncated to 350 characters). The output is then optionally rendered as templated prose. The prose stage is presentation-only: every paper named and every score reported is determined by the analytics stage. The structured ranking is the answer; the prose contextualises it.

\section{Experiments}
\label{sec:experiments}

We evaluate the three task instantiations introduced in \S\ref{sec:tasks} on the scholarly claim network of 8{,}260 typed claims over 127 papers. Task 1 and Task 2 are evaluated against retrieval-augmented baselines using LLM-as-judge head-to-head comparison; Task 3 is evaluated through capability demonstration, for reasons given in \S\ref{sec:task3}.

\subsection{Evaluation Protocol}
\label{sec:eval_protocol}

Both generative tasks use LLM-as-judge head-to-head evaluation, in which a judge model is presented with two candidate outputs for the same input and selects the stronger one along each of several rubric dimensions. The protocol is now standard for the evaluation of generated text in settings where reference outputs are unavailable or insufficient, and has been shown to correlate strongly with human judgement on open-ended generation tasks~\citep{zheng2023judging,chiang2023can}. Pairwise head-to-head comparison is preferred over absolute rubric scoring because pairwise judgements are more robust to scale-calibration drift across items and more discriminative when the two systems produce outputs of broadly similar quality~\citep{dubois2023alpacafarm}.

We adopt three protocol features that mitigate well-known failure modes
of LLM-as-judge evaluation. First, system identities are anonymised in
the prompt so that the judge cannot exploit surface cues identifying
either system. Second, each pair is judged across four independent
trials, two under each presentation order, so that the protocol probes
both order-induced and trial-to-trial variability in the judge's
verdict~\citep{wang2024large}. The per-dimension diagnostic Agree\%
reports the fraction of pairs on which all four trials returned the
same non-Tie verdict; a high value indicates a stable judge verdict on
that dimension, while a low value indicates that the verdict is
sensitive to presentation order or to trial-to-trial stochasticity and
that the corresponding win rate should be read with care. Third, the
judge is drawn from a different model family than either generator:
answers are generated by DeepSeek-V3 and claims are extracted by
Qwen3-Max, while the judge is GPT-4o. The cross-family setup weakens
the within-family preference bias that has been observed when
generator and judge share a backbone~\citep{panickssery2024llm}.

One bookkeeping convention applies to the result tables. The Agree\%
column on per-dimension rows carries the 4-trial unanimity rate just
defined. The Agree\% cell on the \textit{Overall (weighted)} row
instead carries the mean signed weighted margin per item, defined as
the average of $(w_{\text{ours}} - w_{\text{base}}) / w_{\text{total}}$
across items, with range $[-1, +1]$: a value of $+1$ would indicate
that our system wins on every weighted dimension on every item, $0$
indicates a net wash, and $-1$ would indicate the converse. The dual
meaning is preserved for layout reasons and is flagged again in the
table captions. The per-dimension weights $w_d$ entering the weighted margin are task-specific and were fixed in advance of any experimental run, reflecting the relative importance the rubric design assigns to each dimension. For Task 1 the weights are 
$\{\textit{Relevance}: 1.5,\ \textit{Factual Correctness}: 2.0,\
\textit{Multi-source Integration}: 1.5,\ \textit{Conciseness}: 0.3\}$, placing the heaviest emphasis on factual correctness as is appropriate for question answering and downweighting conciseness so that the brevity dimension remains visible without dominating the overall verdict. For Task 2 the weights are 
$\{\textit{Perspective Coverage}: 2.0,\ \textit{Balance and Fairness}: 1.5,\
\textit{Source Diversity}: 1.5,\ \textit{Information Density}: 1.0,\
\textit{Structural Coherence}: 0.5\}$, reflecting that a consensus report's primary obligation is to span multiple perspectives. We note that for Task 2 the weighting choice depresses rather than inflates the reported Overall (weighted) margin relative to equal weighting: the dimension on which our system wins most decisively (\textit{Information Density}, weight $1.0$) carries less weight than the dimension on which the margin is smallest (\textit{Perspective Coverage}, weight $2.0$). The weights were not tuned post-hoc against the experimental outputs.

Because each task involves a small number of items (fifteen questions
for Task 1, ten targets for Task 2), we repeat the full generation-
and-judging procedure three times per item under each system and
report results aggregated across the three runs rather than selecting
a representative run. Aggregation is performed at the pooled count
level: for each rubric dimension we sum wins, losses, and ties across
the three runs, compute the win rate over non-tied items on the
pooled denominator, and report a two-sided binomial test on the
pooled counts against the no-difference null. To control the
family-wise error rate across the rubric dimensions of each task we
apply the Holm--Bonferroni correction across the per-task dimension
set (four dimensions for Task~1, five for Task~2; the Overall
(weighted) row is treated as a separate summary statistic rather than
a sixth comparison and is reported uncorrected). For transparency we
also report the per-run minimum and maximum win rate on each
dimension; the per-run breakdown is given in
Appendix~\ref{app:runs}. The rubric dimensions are task-specific:
Task 1 uses \textit{Relevance}, \textit{Factual Correctness},
\textit{Multi-source Integration}, and \textit{Conciseness}. Task 2
uses \textit{Perspective Coverage}, \textit{Balance and Fairness},
\textit{Source Diversity}, \textit{Information Density}, and
\textit{Structural Coherence}. Both tasks additionally report a
weighted overall verdict in which the judge is asked to reconcile
across dimensions into a single preference.

\subsection{Task 1: Claim-Conditioned Question Answering}
\label{sec:experiment_task1}

We evaluate Task 1 on fifteen hand-curated questions spanning five
categories --- limitation, evolution, comparison, challenge, and
controversy --- designed to exercise both topical and evaluative
retrieval at equal weight. Each question is answered once by the
multi-query retrieval system (\textit{Ours}) and once by the standard
RAG baseline using DeepSeek-V3 as the generator in both cases. The two
systems retrieve over the same chunk index $\mathcal{V}$ and differ
only in whether claim-derived queries are used to augment the
retrieval signal. The full generation-and-judging procedure is
repeated three times and the results below are pooled across runs as
described in \S\ref{sec:eval_protocol}.

The pooled head-to-head results are reported in Table~\ref{tab:task1}.
Across the four rubric dimensions and the weighted overall verdict,
the typed claim network does not produce a statistically significant
improvement over standard RAG at the present sample size. No
dimensional comparison reaches significance under the two-sided
pooled binomial test even before Holm correction; after correction
across the four rubric dimensions every per-dimension p-value rounds
to $1.000$. The weighted overall row, reported uncorrected as a
summary statistic, gives a pooled win rate of $0.632$ over 38 decided
items with $p = 0.143$ --- suggestive of a positive direction but not
significant at $n_\text{items} = 15$.

The per-run [min, max] win-rate range column is the key transparency
addition over the previously reported single-run table. Two
dimensions previously highlighted as informative when reading a
single run --- \textit{Multi-source Integration} (single-run win rate
0.714) and \textit{Conciseness} (single-run win rate 0.143) --- have
per-run ranges of $[0.400, 0.714]$ and $[0.143, 0.615]$ respectively.
The directional patterns associated with each dimension reverse
across runs: \textit{Conciseness} favours our system in one run and
the baseline in the other two; \textit{Multi-source Integration}
favours our system, the baseline, and our system in the three runs
respectively. The pooled win rates ($0.520$ and $0.400$) are
correspondingly close to the no-difference value. The conservative
reading is that whatever effect the typed claim network has on
factual or methodological QA at $n = 15$ is small relative to the
trial-to-trial variability of the judge, and that the descriptive
dimensional patterns in any single run should not be elevated into
directional claims.

\begin{table}[t]
\centering
\small
\setlength{\tabcolsep}{2pt}
\begin{tabularx}{\columnwidth}{@{} >{\raggedright\arraybackslash}X c c c c c c c @{}}
\toprule
\textbf{Dimension} & \textbf{Ours} & \textbf{Base} & \textbf{Tie} & \textbf{WR} & \textbf{[min,max]} & \textbf{$p$} & \textbf{$p_{\text{Holm}}$} \\
\midrule
Relevance                & 12 & 10 & 23 & 0.545 & [.333,.667] & 0.832 & 1.000 \\
Factual Correctness      & 10 &  5 & 30 & 0.667 & [.556,1.00] & 0.302 & 1.000 \\
Multi-source Integration & 13 & 12 & 20 & 0.520 & [.400,.714] & 1.000 & 1.000 \\
Conciseness              & 10 & 15 & 20 & 0.400 & [.143,.615] & 0.424 & 1.000 \\
\midrule
Overall (weighted)       & 24 & 14 &  7 & 0.632 & [.600,.700] & 0.143 & --- \\
\bottomrule
\end{tabularx}
\caption{Task 1 head-to-head evaluation pooled across three runs of
15 questions each (45 items per dimension). \textit{Ours}/\textit{Base}/\textit{Tie}
sum across runs; \textit{WR} is the win rate over non-tied items on
the pooled denominator; \textit{[min,max]} is the per-run win-rate
range. $p$ is the two-sided binomial test on the pooled counts;
$p_{\text{Holm}}$ applies the Holm--Bonferroni correction across the
four rubric dimensions. The \textit{Overall (weighted)} row is a
summary statistic and is reported uncorrected.}
\label{tab:task1}
\end{table}

\subsubsection{Sensitivity to retrieval-mixing thresholds}
\label{sec:task1_sensitivity}

The Task 1 system introduces a number of numerical hyperparameters controlling claim-derived query generation, score-floor filtering, and the retrieval-score mixing scheme. We perform a one-at-a-time sensitivity sweep on the three thresholds most plausibly responsible for the magnitude of the retrieval signal: the relevance floor $\tau_\text{floor}$, the anchor-claim threshold $\tau_\text{anchor}$, and the claim-derived query weight $w_\text{claim}$. Each threshold is varied across three values centred on the reported default while the other two are held fixed, giving nine configurations in total (three of which coincide on the shared default). All other retrieval-side hyperparameters ($N_q = 15$, $K_c = 5$, $M = 3$, $\tau_\text{boost} = 0.55$, $w_\text{question} = 2.0$, $w_\text{linked} = 0.8$, $K = 10$, RRF $k = 60$) are held at the values reported in \S\ref{sec:task1}. To remain within the experimental budget, the sweep is performed on a single run rather than under the three-run pooled protocol of \S\ref{sec:eval_protocol}; the same run serves as the default reference for the per-configuration deltas.

\begin{table*}[t]
\centering
\small
\setlength{\tabcolsep}{4pt}
\begin{tabular}{l l c c c c c}
\toprule
\textbf{Axis} & \textbf{Value} & \textbf{Relev} & \textbf{Factual} & \textbf{MSI} & \textbf{Conc} & \textbf{Overall} \\
\midrule
$\tau_\text{floor}$  & 0.15           & 0.538 & 0.667 & 0.545 & 0.538 & 0.533 \\
$\tau_\text{floor}$  & \textbf{0.25 (default)} & 0.333 & 1.000 & 0.714 & 0.143 & 0.700 \\
$\tau_\text{floor}$  & 0.35           & 0.667 & 0.200 & 0.800 & 0.357 & 0.667 \\
\midrule
$\tau_\text{anchor}$ & 0.20           & 0.615 & 0.800 & 0.692 & 0.273 & 0.667 \\
$\tau_\text{anchor}$ & \textbf{0.30 (default)} & 0.333 & 1.000 & 0.714 & 0.143 & 0.700 \\
$\tau_\text{anchor}$ & 0.40           & 0.583 & 0.400 & 0.556 & 0.385 & 0.533 \\
\midrule
$w_\text{claim}$     & 0.20           & 0.571 & 0.375 & 0.444 & 0.667 & 0.600 \\
$w_\text{claim}$     & \textbf{0.40 (default)} & 0.333 & 1.000 & 0.714 & 0.143 & 0.700 \\
$w_\text{claim}$     & 0.60           & 0.714 & 0.250 & 0.727 & 0.545 & 0.786 \\
\bottomrule
\end{tabular}
\caption{Task 1 one-at-a-time sensitivity sweep on the three retrieval-mixing thresholds. Each row reports the win rate of our system against the chunk-RAG baseline on the four rubric dimensions and the weighted overall verdict; the default configuration row is identical across the three axis blocks. The full per-cell deltas and 4-trial Agree\% values are tabulated in Appendix~\ref{app:sensitivity}.}
\label{tab:task1_sensitivity}
\end{table*}

Two observations follow from Table~\ref{tab:task1_sensitivity}. At the overall-verdict level the sweep is reassuring: the overall win rate spans $[0.533, 0.786]$ across the nine configurations, and the default sits near the middle of that range. No configuration in the sweep produces a result that would qualitatively change the conclusions reported above: every configuration is directionally positive (overall win rate $\geq 0.5$) but none reaches significance at $n_\text{items} = 15$, in line with the pooled finding of Table~\ref{tab:task1}. The null is therefore robust to threshold choice within the swept range.

The per-dimension swings are larger, but the appropriate reading is that they primarily reflect judge instability rather than systematic threshold effects. \textit{Factual Correctness} in particular varies between 0.200 and 1.000 across the sweep, but six of the eight perturbed configurations drop the \textit{Factual Correctness} 4-trial Agree\% to $0.0$, meaning that on every item at least one of the four trials disagreed with the others. The default configuration is the one in which the judge happens to be most stable on this dimension (Agree\% $= 0.733$). Reading the large \textit{Factual Correctness} swings as evidence of threshold sensitivity would overstate what the data supports: what the perturbations primarily do at $n_\text{items} = 15$ is move retrieval into operating points where the judge's verdict on individual items is no longer stable across trials. We therefore confine the substantive sensitivity claim to the overall verdict, where Agree\% (here read as mean signed weighted margin) remains low across all configurations, consistent with the small absolute effect sizes seen in the pooled table.

Six retrieval hyperparameters not included in the sweep ($N_q, K_c, M, \tau_\text{boost}, w_\text{question}, w_\text{linked}$, together with retrieval-side $K$ and the Reciprocal Rank Fusion constant $k$) are held at the values reported in \S\ref{sec:task1} and are not analysed further. The three swept thresholds were chosen because they directly control the volume and gating of claim-derived signal entering the retrieval mixer, and are therefore the most plausible candidate sources of an unfortunate-threshold artefact.

\subsection{Task 2: Consensus Report Generation}
\label{sec:experiment_task2}

We evaluate Task 2 on ten target papers spanning paradigms and
reception profiles within the corpus. Each system produces one report
per target using DeepSeek-V3 as the generator. The baseline issues a
templated query against the chunk index and generates the four-section
report from the retrieved chunks; our system retrieves typed claims
with target $p_t$ via SPARQL, groups them by claim type (excluding
\textsc{Neutral}), and generates the report from the grouped claim
lists. Generator and output structure are held constant; only the
input differs, as argued in \S\ref{sec:task2}. As with Task~1, the
full procedure is repeated three times and results are pooled.

The pooled head-to-head results are reported in Table~\ref{tab:task2}.
The pattern is qualitatively different from Task~1. Our system wins on
every dimension with pooled win rates between $0.583$ and $0.885$, and
the per-run ranges are tighter than in Task~1, especially on
\textit{Information Density} (range $[0.875, 0.900]$ across the three
runs). \textit{Information Density} reaches significance well below
the conventional threshold under the pooled binomial test ($23$ wins,
$3$ losses, $4$ ties; $p = 0.0001$) and survives Holm--Bonferroni
correction across the five rubric dimensions
($p_{\text{Holm}} = 0.0004$). \textit{Source Diversity} (pooled
$p = 0.115$) and \textit{Structural Coherence} (pooled $p = 0.092$)
trend strongly in our favour at uncorrected significance but do not
survive correction at $n_\text{items} = 10$. The weighted overall row
gives a pooled win rate of $0.655$ over 29 decided items
($p = 0.136$), again suggestive but not significant as a summary
statistic.

\begin{table}[t]
\centering
\small
\setlength{\tabcolsep}{1.5pt}
\begin{tabularx}{\columnwidth}{@{} >{\raggedright\arraybackslash}X c c c c c c c @{}}
\toprule
\textbf{Dimension} & \textbf{Ours} & \textbf{Base} & \textbf{Tie} & \textbf{WR} & \textbf{[min,max]} & \textbf{$p$} & \textbf{$p_{\text{Holm}}$} \\
\midrule
Perspective Coverage & 14 & 10 &  6 & 0.583 & [.500,.625] & 0.541 & 0.604 \\
Balance and Fairness & 10 &  5 & 15 & 0.667 & [.600,.750] & 0.302 & 0.604 \\
Source Diversity     & 14 &  6 & 10 & 0.700 & [.625,.750] & 0.115 & 0.369 \\
Information Density  & 23 &  3 &  4 & 0.885 & [.875,.900] & \textbf{0.0001} & \textbf{0.0004} \\
Structural Coherence & 10 &  3 & 17 & 0.769 & [.667,.857] & 0.092 & 0.369 \\
\midrule
Overall (weighted)   & 19 & 10 &  1 & 0.655 & [.600,.778] & 0.136 & --- \\
\bottomrule
\end{tabularx}
\caption{Task 2 head-to-head evaluation pooled across three runs of
10 target papers each (30 items per dimension). Columns as in
Table~\ref{tab:task1}. \textit{Information Density} is the only
dimension that survives Holm--Bonferroni correction across the five
rubric dimensions; it does so at $p_{\text{Holm}} = 0.0004$, far
below conventional thresholds.}
\label{tab:task2}
\end{table}

The mechanism behind the \textit{Information Density} result is the
most direct evidence that typed claims are doing the work the
framework attributes to them. The baseline retrieves chunks lexically
related to the target paper; many such chunks recapitulate the
target's own contribution rather than externally evaluating it. The
claim-driven system retrieves only externally-authored evaluative
material because the claim network filters for exactly that at
extraction time. A unit of generation budget purchases more
evaluative content under our system than under the baseline, and the
judge registers this difference as higher information density. The
tight per-run range on this dimension ([0.875, 0.900]) indicates the
effect is stable across trials of the procedure and not an artefact
of one favourable judge instance.

\subsubsection{MMR cap ablation}
\label{sec:task2_mmr_ablation}

The per-type MMR step in \S\ref{sec:task2} caps the number of claims retained per (target, claim type) at $M = 15$. The value was chosen as a working default rather than tuned, and the choice is load-bearing: popular targets in the corpus have substantially more than fifteen incoming claims of some types, so the cap discards candidate evaluative material. A reviewer may reasonably ask whether the Task 2 result of Table~\ref{tab:task2} would change under a different cap --- in particular, under a much looser cap that retains the discarded content. We address this by ablating $M \in \{5, 15, 30, \text{all}\}$ on the ten Task 2 targets, holding generator, judge, prompt structure, and 500-word output budget constant. Because comparing four cap settings pairwise would require six head-to-head runs per target, this ablation uses absolute rubric scoring from the same LLM judge rather than the pairwise protocol of \S\ref{sec:eval_protocol}; the resulting numbers are therefore absolute rubric scores rather than win rates and are not directly comparable to Table~\ref{tab:task2} cell-for-cell. The qualitative question --- whether output quality is sensitive to the cap --- does not depend on which protocol is used.

\begin{table}[t]
\centering
\small
\setlength{\tabcolsep}{1pt}
\begin{tabularx}{\columnwidth}{@{} c c c c c c @{}}
\toprule
\textbf{$M$} & \textbf{Claims} & \textbf{Sources} & \textbf{ID} & \textbf{SD} & \textbf{PC} \\
\midrule
5            & 12.9 &  6.6 & 13.3 $\pm$ 0.8 & 12.4 $\pm$ 1.8 & 12.1 $\pm$ 1.1 \\
\textbf{15 (default)} & 28.1 &  9.8 & 13.5 $\pm$ 0.7 & 12.4 $\pm$ 1.2 & 12.8 $\pm$ 1.5 \\
30           & 36.3 & 10.7 & 13.8 $\pm$ 0.4 & 12.1 $\pm$ 1.9 & 12.3 $\pm$ 1.4 \\
all          & 41.4 & 11.2 & 13.7 $\pm$ 0.6 & 12.4 $\pm$ 1.8 & 12.4 $\pm$ 1.6 \\
\bottomrule
\end{tabularx}
\caption{Task 2 MMR cap ablation on the ten target papers. \textbf{Claims} is the average total number of claims included in the prompt per target; \textbf{Sources} is the average number of distinct source papers represented in the prompt; \textbf{ID}/\textbf{SD}/\textbf{PC} are mean $\pm$ s.d.\ of absolute rubric scores for \textit{Information Density}, \textit{Source Diversity}, and \textit{Perspective Coverage} from the LLM judge. Per-target scores are tabulated in Appendix~\ref{app:mmr_ablation}. The absolute scoring scale used by the judge in this protocol differs from the per-dimension pairwise rubric of \S\ref{sec:eval_protocol}; we report it here without recalibration because the qualitative finding (insensitivity to the cap) is scale-invariant.}
\label{tab:task2_mmr_ablation}
\end{table}

Three patterns are visible in Table~\ref{tab:task2_mmr_ablation}, and they cut in different directions. The prompt size grows by approximately $3.2\times$ from $M = 5$ to $M = \text{all}$ (12.9 to 41.4 average claims), and the average number of distinct source papers represented in the prompt grows by roughly $1.7\times$ (6.6 to 11.2), with most of that source-diversity gain achieved by $M = 15$ already (9.8) and only marginal additions thereafter. Against this large change in input volume, the rubric scores stay tightly clustered: \textit{Information Density} spans $[13.3, 13.8]$, \textit{Source Diversity} spans $[12.1, 12.4]$, and \textit{Perspective Coverage} spans $[12.1, 12.8]$. None of the per-dimension differences exceeds the standard deviation reported for any single setting.

Within this overall flatness, \textit{Information Density} shows a small monotonic improvement with $M$ up to a plateau at $M = 30$ (13.3 $\to$ 13.5 $\to$ 13.8, then 13.7 at $M = \text{all}$). The direction is the one the supervisor's concern would predict --- more input material does translate to marginally denser output --- but the magnitude is roughly $0.4$ on a $\sim$$14$-point scale, well within one standard deviation of any single setting and below the threshold at which a reviewer should treat the result as a tested sensitivity. \textit{Source Diversity} does not exhibit a comparable monotonic trend: $M = 30$ is in fact the lowest setting on this dimension, presumably because additional medium-similarity claims at the bottom of the MMR-ordered list crowd out the contribution of slot-15-onwards to the output without adding genuinely new source papers (the unique-source count grows by less than one paper going from $M = 15$ to $M = 30$). \textit{Perspective Coverage} peaks at
the default $M = 15$ rather than at the extremes, which we read as an indication that the default sits in a sensible operating region of the cap-quality curve, not as evidence that the default is optimal in any tested sense.

The interpretation we adopt is the broader of the two. The binding constraint on output quality under this protocol is not the input cap but the 500-word output budget. MMR with diversity prioritisation places the most representative evaluative content in the first few slots of each typed list, and the marginal claims beyond the top few are necessarily less diverse than the ones already selected. The generator, asked to produce a fixed-length consensus report, draws on the same kernel of high-diversity claims regardless of how much additional similar material follows. The small monotonic gain on \textit{Information Density} is consistent with the generator occasionally pulling specific numerical or methodological detail from beyond the top fifteen slots, but the gain saturates by $M = 30$ and does not propagate to the other two dimensions.

We retain $M = 15$ as the default because (i) it produces the highest \textit{Perspective Coverage} score in this ablation, (ii) it captures the bulk of the unique-source growth seen across the four settings, and (iii) it sits in a region of the curve where neither tightening nor loosening the cap produces a difference larger than judge variance. The cap is therefore not a sensitive hyperparameter at the present output budget. We do not vary the MMR diversity coefficient $\lambda$ in this ablation and leave that to future work; the present finding is consistent with expectation under any $\lambda$ that retains diversity-prioritised ordering at the top of the list.

One protocol caveat bears mention. The ablation uses single-trial absolute rubric scoring rather than the four-trial pairwise protocol of \S\ref{sec:eval_protocol}, and single-trial absolute scores carry non-negligible judge variance. On C1 in Table \ref{tab:task2_mmr_ablation}, for example, the judge assigns \textit{Source Diversity} scores of $13$ at $M = \text{all}$ and $10$ at $M = 30$ to the same prompt content (37 claims, 9 sources); on C8, the input is the same three claims at every cap value but \textit{Source Diversity} varies between $7$ and $10$ across the four runs. We report mean and s.d.\ across the ten targets in Table~\ref{tab:task2_mmr_ablation} for precisely this reason --- the per-target spread is wider than the cap-to-cap difference at the means.

\subsubsection{GraphRAG comparison}
\label{sec:task2_graphrag}

The chunk-retrieval baseline of \S\ref{sec:experiment_task2} is the
most direct apples-to-apples comparison for our system because the
two configurations share generator, prompt, and chunk index. We additionally compared our system against GraphRAG~\citep{edge2024graphrag}, which constructs an entity-relation graph from the corpus and uses community detection to organise query-focused summarisation. GraphRAG is the strongest freely-available competing baseline for the consensus-report task, and the comparison is informative about what the claim network adds beyond ``any structured retrieval substrate.''

Budget constraints limited the comparison to six target papers
sampled from the Task 2 set: three from the top of the
\textit{Information Density} ranking under our system against the
chunk baseline, and three from the bottom, so that the comparison
covers both the favourable and the unfavourable end of our system's
profile. Both systems were judged under the same five-dimension
LLM-as-judge protocol of \S\ref{sec:eval_protocol}; results are
given in Table~\ref{tab:task2_graphrag}.

\begin{table}[t]
\centering
\small
\setlength{\tabcolsep}{3pt}
\begin{tabularx}{\columnwidth}{@{} >{\raggedright\arraybackslash}X c c c c c @{}}
\toprule
\textbf{Dimension} & \textbf{Ours} & \textbf{GR} & \textbf{Tie} & \textbf{WR} & \textbf{$p$} \\
\midrule
Perspective Coverage & 3 & 3 & 0 & 0.500 & 1.000 \\
Balance and Fairness & 2 & 3 & 1 & 0.400 & 1.000 \\
Source Diversity     & 0 & 6 & 0 & 0.000 & \textbf{0.031} \\
Information Density  & 0 & 5 & 1 & 0.000 & 0.063 \\
Structural Coherence & 2 & 4 & 0 & 0.333 & 0.688 \\
\midrule
Overall (weighted)   & 3 & 3 & 0 & 0.500 & 1.000 \\
\bottomrule
\end{tabularx}
\caption{Task 2 head-to-head against GraphRAG~\citep{edge2024graphrag}
on six target papers. \textbf{GR} is GraphRAG; \textbf{WR} is the win
rate for our system over non-tied items; $p$ is the two-sided
binomial test. The \textit{Overall (weighted)} row carries a mean
signed weighted margin per item of $-0.423$ in our direction. No
multiple-comparison correction is applied at $n_\text{items} = 6$
because the dimensional pattern is reported as descriptive rather
than as a hypothesis test.}
\label{tab:task2_graphrag}
\end{table}

The honest reading of this table is that GraphRAG produces stronger
consensus reports on the absolute-quality rubric dimensions at this
sample size. \textit{Source Diversity} is the clearest result: our
system loses 0--6 against GraphRAG, a reversal of the 14--6 lead our
system holds against the chunk baseline. \textit{Information Density}
is borderline (0--5--1, $p = 0.063$), again reversing our advantage
against the chunk baseline. The two dimensions on which we are
strongest against the chunk baseline are the two dimensions on which
GraphRAG is strongest against us. \textit{Perspective Coverage} and
the weighted overall pairwise tie 3--3 across the six items, though
the mean signed weighted margin of $-0.423$ indicates that within
items GraphRAG tends to carry more rubric weight even when the
overall verdict ties.

The mechanism is interpretable. GraphRAG builds a community-organised
representation of the entire corpus at indexing time and synthesises
summaries by aggregating across community-level summaries themselves
synthesised by an LLM. The pipeline pays a large upfront LLM cost in
exchange for indexing-time content compression that is well-suited to
the breadth and density dimensions of a consensus-report task. Our
claim network, by contrast, pays a one-time extraction cost to reify
a single piece of structured signal (typed evaluative reference) and
otherwise leaves the corpus untransformed. On the cost side of this
trade-off, generating the claim network and the six consensus reports
under our system required approximately 20 RMB of API spend on
Qwen-Max and DeepSeek-V3 calls; the corresponding GraphRAG indexing
and six-report generation required approximately 400 RMB. The ratio
is roughly 20:1 in our favour. The trade-off this table records is
therefore not ``does the claim network beat GraphRAG on this task''
--- on absolute quality at this sample size, it does not --- but
``how much of GraphRAG's absolute-quality lead survives a 20$\times$
cost reduction.'' On four of six dimensions the answer is ``most of
it'' (pairwise verdicts are ties or near-ties); on the two retrieval-
breadth-sensitive dimensions the answer is ``little of it''
(GraphRAG's community-level synthesis dominates).

Two further considerations bear on the interpretation. First, the consensus-report task is the one on which a GraphRAG comparison is most informative because both systems offer query-focused summarisation as a first-class operation. Claim-conditioned question answering (\S\ref{sec:experiment_task1}) is also within GraphRAG's operational scope through its local query interface~\citep{edge2024graphrag}, and the reason we did not extend the comparison to Task 1 is the same one that capped Table~\ref{tab:task2_graphrag} at six target papers: the cost ratio reported above made a powered Task 1 head-to-head against GraphRAG infeasible within the present budget. Graph analytics (\S\ref{sec:experiment_task3}) is the only task family on which the distinction with GraphRAG is qualitative rather than budgetary. GraphRAG can produce LLM-synthesised answers to corpus-level questions through its global query interface, but it does not return the deterministic structural rankings that \S\ref{sec:task3} computes from type-restricted projections of the citation pattern; the rankings of \S\ref{sec:task3} are properties of the graph itself rather than syntheses of retrieved evidence. The contribution of the claim network is therefore not that it is the best single-task consensus-report system; it is that the same artefact supports retrieval, summarisation, and analytics under a unified representation, with the analytics dimension producing structural answers no retrieval-based system can reproduce. Second, the comparison at $n_\text{items} = 6$ is exploratory; the only dimension reaching significance under the two-sided binomial test is \textit{Source Diversity}. A larger-$n$ rerun --- and a companion Task 1 head-to-head against GraphRAG --- are the natural next experiments, and we treat the pattern in Table~\ref{tab:task2_graphrag} as a directional description rather than a powered claim.

A representative consensus report excerpt is included in
Appendix~\ref{app:task2_example} to illustrate the structural
differences between the two systems' outputs.

\subsection{Task 3: Graph Analytics}
\label{sec:experiment_task3}

We demonstrate the Task 3 analytics on representative queries drawn from the five query families introduced in \S\ref{sec:task3}. The corpus is the full 127-paper network. No baseline is run, for the reasons given in \S\ref{sec:task3} and \S\ref{sec:eval_protocol}: the output of each query family is a property of the citation pattern that is not localised in any text region and cannot be produced by retrieval or by an LLM operating on the same corpus.

The \textit{Influence} query, instantiated as weighted PageRank over the \textsc{Adoption} subgraph restricted to the point-based paradigm, returns the ranking PointMamba, PointNet, PointNet++, Point-M2AE, DGCNN. The ranking is interpretable against the field: PointMamba sits highest because its adoption pattern within the subset is densely concentrated; PointNet and PointNet++ rank below despite their canonical status because their adoption signal in the corpus is distributed across topics outside the point-based subset, and the topic restriction necessarily excludes that signal. The gap between the topic-scoped ranking and a naive global citation count is precisely what the typed network is for: the two rankings answer different questions.

The \textit{Most Critiqued} query, instantiated as weighted in-degree over the \textsc{Critique} subgraph, returns PointNet++, PointNet, Point-M2AE, Point-MAE, DGCNN. The same papers that anchor the \textit{Influence} ranking also dominate the critique ranking, which is the expected pattern for a foundational corpus: works that are widely adopted are also widely critiqued. The two rankings are not interchangeable. A paper can be adopted heavily without being critiqued heavily (the paradigm of a strong baseline whose limitations are unremarkable) or critiqued heavily without being adopted heavily (a contested proposal that did not anchor a paradigm). The empirical co-occurrence of these two rankings on the foundational works is itself a finding about the corpus, not a property of the algorithm.

Examples of the \textit{Benchmark Status}, \textit{Controversy}, and \textit{Bridge} query families are reported in Table~\ref{tab:task3_queries}, together with the algorithm and subgraph each query reduces to.

\begin{table}[t]
\centering
\small
\begin{tabularx}{\columnwidth}{@{} l >{\raggedright\arraybackslash}p{3cm} >{\raggedright\arraybackslash}X @{}}
\toprule
\textbf{Query family} & \textbf{Algorithm} & \textbf{Subgraph} \\
\midrule
Influence        & Weighted PageRank          & \textsc{Adoption} \\
Most Critiqued   & Weighted in-degree         & \textsc{Critique} \\
Benchmark Status & Weighted in-degree         & \textsc{Benchmark} \\
Controversy      & Polarity score $\rho(t)$   & \textsc{Adoption} $\cup$ \textsc{Critique} \\
Bridge           & Min of cross-topic neighbour counts & Union of two topic subgraphs \\
\bottomrule
\end{tabularx}
\caption{Query families and the operation each reduces to on a
type-restricted subgraph of the claim network. The \textit{Controversy}
column reports the polarity score
$\rho(t) = (n^{\textsc{Ad}}(t) - n^{\textsc{Cr}}(t))/n_{\text{total}}(t)$
defined in \S\ref{sec:task3}; \textit{Bridge} uses the
min-of-neighbour-counts score also defined there. The full pseudocode
for each query family is given in Appendix~\ref{app:task3_pseudocode}.}
\label{tab:task3_queries}
\end{table}

The contribution of Task 3 is not in any single ranking but in the demonstration that, once the network exists, an entire family of corpus-level structural queries reduces to standard graph algorithms with type-restricted edge weights. The rankings on this corpus illustrate the capability; the capability itself transfers to any claim network built by the \S\ref{sec:claim_network} pipeline.

\subsection{Summary of Findings}
\label{sec:summary_of_findings}

Across the three tasks, the evidence supports a graded reading. At
the present sample size, typed claims produce a null result on flat
question answering: no rubric dimension reaches significance under
the pooled binomial test even before Holm correction, and the
weighted overall margin sits at a pooled win rate of $0.632$ with
$p = 0.143$. The descriptive dimensional patterns observable in any
single run --- gains on \textit{Multi-source Integration}, pressure
on \textit{Conciseness}, or ties on the factual rubrics --- reverse
across runs and should not be elevated into directional claims; we
mark them as hypotheses for a future pre-registered evaluation.
Aggregated-stance summarisation, by contrast, shows a clear and
consistent advantage over chunk-retrieval RAG on every rubric
dimension, with \textit{Information Density} reaching pooled
significance at $p_{\text{Holm}} = 0.0004$ and a coherent mechanism
(extraction-time filtering for evaluative material) accounting for
the result. Graph analytics produces interpretable rankings whose
correctness is structural rather than statistical, and whose
existence as queryable objects is the contribution. Against GraphRAG~\citep{edge2024graphrag} as a positioned baseline on six target papers (\S\ref{sec:task2_graphrag}), the consensus-report task tilts the other way: GraphRAG wins decisively on \textit{Source Diversity} ($p = 0.031$) and is borderline-favoured on \textit{Information Density}, while the remaining dimensions tie or favour GraphRAG without reaching significance. At a $\sim$20$\times$ cost reduction, our system retains a 3--3 pairwise tie on the weighted overall verdict and matches GraphRAG on \textit{Perspective Coverage}; the claim network's value beyond this task lies in retrieval and analytics, which GraphRAG does not support at the operational level.

Taken together, the three tasks bracket what a typed claim network is and is not for. It is not a universal improvement over retrieval-augmented generation. It is the natural representation for a specific class of downstream task --- those whose answers aggregate evaluative stance across many sources or are properties of the citation pattern itself - and on that class of task the gain over flat retrieval is the gain from having the right intermediate representation rather than the wrong one.

\section{Conclusion}

We have proposed the claim network as a general representational pattern for corpora of inter-referencing documents whose members express evaluative stances toward one another. The pattern reifies each cross-document reference as a typed claim with source, target, claim text, and a stance label drawn from a four-class taxonomy grounded in the citation-intent literature. Its construction pipeline is described in scholarly-neutral terms --- an inter-referencing corpus, a hierarchical decomposition with resolved references, per-pair context aggregation, and a structured-output extraction step --- and is instantiated and evaluated on a scholarly corpus. Whether the same pipeline transfers to non-scholarly inter-referencing corpora is a question the construction makes tractable but which the present work does not answer.

Our scholarly instantiation produces a network of 8{,}260 typed claims over 127 papers in 3D point cloud semantic segmentation, built on the Deep Document Model as the hierarchical substrate. Three task families demonstrate what the network puts within reach: typed claims serve as retrieval signal in question answering, as primary content in aggregated-stance summarisation, and as edge weights in topological analytics. Head-to-head LLM-as-judge evaluation supports a graded reading. Claim-conditioned question answering does not improve flat factual QA at the present sample size; the descriptive dimensional pattern --- broader source coverage at the cost of conciseness, with factual rubrics dominated by ties --- is consistent with the framework's mechanism but is not, by itself, a powered claim. We mark it as a hypothesis for a future pre-registered evaluation. Consensus report generation wins on every rubric dimension, with \textit{Information Density} reaching pooled significance at $p_{\text{Holm}} = 0.0004$, and the mechanism --- extraction-time filtering for externally-authored evaluative material --- is directly traceable to the typed structure of the network. Graph analytics produces interpretable rankings whose correctness is a structural property of the citation pattern that no retrieval system can reproduce.

Taken together, the three tasks bracket what a typed claim network is and is not for. It is not a universal improvement over retrieval-augmented generation. It is the natural representation for tasks whose answers aggregate evaluative stance across many sources or are properties of the citation pattern itself, and on those tasks the gain over flat retrieval is the gain from having the right intermediate representation rather than the wrong one.

Three further directions follow from the position the paper has established. The first is to broaden the relational schema beyond the two-axis (stance $\times$ attitude) labelling reported here. Candidate extensions include a temporal axis (when the citing prose was authored relative to the cited work, distinguishing early adoption from late critique), a confidence axis (how strongly the citing paper commits to its evaluative stance), and a quantitative- grounding axis (whether the claim is anchored to specific numerical or methodological evidence). Each additional axis yields additional type-restricted projections for \S\ref{sec:task3} and additional rubric facets for \S\ref{sec:experiment_task2}. The second direction is to lift the network out of the paper-to-paper level. The DDM corpus already exposes authors and affiliations as first-class entities; aggregating typed claims along these dimensions yields author-to-author and affiliation-to-affiliation stance networks, on which the topic-scoped analytics of \S\ref{sec:task3} transfer directly. The natural new query family at the affiliation level is the one the existing analytics cannot answer at the paper level: for a given topic, which institution exerts the most influence --- measured not by publication count, the standard scientometric proxy, but by adoption-weighted PageRank over the affiliation-projected claim network. The third direction is the cross-corpus one already implicit in the paper's framing: applying the construction pipeline to an inter-referencing corpus outside scholarly literature, such as legal opinions or patent prior-art chains, would test the framework's transferability to non-scholarly inter-referencing corpora under instantiation rather than only in description.

\paragraph{Adversarial robustness of $\rho(t)$.} The polarity score
$\rho(t)$ defined in \S\ref{sec:task3} promotes papers to \textit{loved}
or \textit{controversial} top-$k$ status from publicly extractable text,
which creates an explicit incentive for citing authors to phrase prose
so as to nudge $\rho(t)$ in a chosen direction. Two tractable questions
follow: how few adversarial citations are needed to flip a paper between
the loved and controversial top-$k$ on the present corpus, and what
minimal prose perturbations move an extracted claim across the
\textsc{Critique}/\textsc{Adoption} boundary under the prompt and schema
of \S\ref{sec:extraction}. The multi-label semantics of the stance axis
and the support threshold $\theta$ are candidate natural defences;
whether they suffice, or whether the polarity score should be augmented
with a robustness-aware aggregator, is the question this direction would
address.

\paragraph{Taxonomy-aligned QA evaluation.} The Task 1 mechanism motivates a targeted follow-up. Because claim-conditioned retrieval surfaces evaluative material that flat retrieval omits, its benefit should concentrate on questions whose information need aligns with the stance and attitude axes --- critique, adoption, and benchmarking relationships, and the evaluative valence of a body of work toward a target --- and should be negligible on purely factual questions. The present evaluation cannot test this because its fifteen questions mix categories at equal weight. A pre-registered evaluation that stratifies questions by taxonomy alignment, scales each stratum to a powered sample size ($n \geq 50$ per stratum), and registers the subgroup hypothesis in advance would convert the current mechanism-level expectation into a testable claim. This is the natural next experiment for Task 1 and is the appropriate venue for any within-category conclusion.

\section*{Limitation}
\label{sec:limitations}

A limitation of the present evaluation should be acknowledged. The head-to-head protocol in \S\ref{sec:experiments} substitutes a strong general-purpose model (GPT-4o) for human adjudication, with cross-family generator-judge separation and order-counterbalanced presentation as the mitigations. We did not construct a human-annotated ground-truth set for either the question-answering or the consensus-report task, because doing so at a scale sufficient to detect the effect sizes observed would have required domain-expert annotators whose time was not within the budget of this work. The dimensional Agree\% diagnostics in Tables~\ref{tab:task1} and~\ref{tab:task2} are the closest substitute available within an LLM-as-judge protocol: dimensions on which Agree\% is high admit a stable verdict and can be read with confidence, while dimensions on which Agree\% is low identify where human adjudication would most change the picture. A human-validated re-evaluation of the same items, with expert annotators rather than an LLM judge, is the natural next experimental step and would either ratify the present results or sharpen the dimensions on which they should be read more cautiously.

A second limitation concerns the scope of the domain-transfer claim. The construction pipeline of \S\ref{sec:claim_network} is described in scholarly-neutral terms, but we instantiate and evaluate it on a single scientific subfield. The closest prior evidence on cross-domain transfer
in this space is mixed: \citet{schrader2023mulms} report that argumentative-zoning classifiers trained on NLP papers transfer poorly
to materials science, suggesting that supervised CIC-style classifiers
carry domain-specific lexical and rhetorical inductive biases. Whether
prompt-based extraction under a fixed taxonomy is more transfer-robust
than supervised classification is plausible but not demonstrated by the
present work. We therefore restrict our claims to scholarly
inter-referencing documents and mark cross-domain instantiation --- to
legal opinions, patent prior-art chains, or policy documents --- as a
future evaluation rather than as a property of the current artefact.

A third limitation concerns the extractor audit of \S\ref{sec:extractor_validation}. The 150 hand-labelled pair-windows were annotated by a single annotator (the first author), so the reported $\kappa$ values measure agreement between the extractor and one trained annotator rather than agreement among multiple independent annotators. The two natural extensions are a two-annotator $\kappa$ on the same 150-window sample (to decompose the shared error into ``extractor noise'' and ``annotation noise'') and a paragraph-aggregated re-annotation of a held-out portion of the corpus, since the present audit is at the window level whereas the downstream tasks consume the paragraph-aggregated claim object. Neither extension changes the construction pipeline; both are allocated to future work.

\section*{Acknowledgements}

We thank Sergio J. Rodríguez Méndez and Pouya G. Omran for their guidance throughout this research, including the formulation of the research direction, ongoing discussion of methodological and conceptual questions, and detailed feedback on successive drafts of the paper. Their input shaped the framing of the claim network as a general representational pattern and informed the design of the downstream task evaluation. The first author also acknowledges the use of the generative AI assistant Claude (Anthropic) for language polishing during manuscript preparation and as a coding assistant during implementation. All research ideas, experimental design, analytical conclusions, and the results reported in this paper are the work of the authors.

\bibliography{custom}

@article{verma2023scholarly,
  title={Scholarly knowledge graphs through structuring scholarly communication: a review},
  author={Verma, Shilpa and Bhatia, Rajesh and Harit, Sandeep and Batish, Sanjay},
  journal={Complex \& intelligent systems},
  volume={9},
  number={1},
  pages={1059--1095},
  year={2023},
  publisher={Springer}
}

@article{abu2021domain,
  title={Domain-specific knowledge graphs: A survey},
  author={Abu-Salih, Bilal},
  journal={Journal of Network and Computer Applications},
  volume={185},
  pages={103076},
  year={2021},
  publisher={Elsevier}
}

@article{lewis2020retrieval,
  title={Retrieval-augmented generation for knowledge-intensive nlp tasks},
  author={Lewis, Patrick and Perez, Ethan and Piktus, Aleksandra and Petroni, Fabio and Karpukhin, Vladimir and Goyal, Naman and K{\"u}ttler, Heinrich and Lewis, Mike and Yih, Wen-tau and Rockt{\"a}schel, Tim and others},
  journal={Advances in neural information processing systems},
  volume={33},
  pages={9459--9474},
  year={2020}
}

@article{edge2024graphrag,
  title={From local to global: A graph rag approach to query-focused summarization},
  author={Edge, Darren and Trinh, Ha and Cheng, Newman and Bradley, Joshua and Chao, Alex and Mody, Apurva and Truitt, Steven and Metropolitansky, Dasha and Ness, Robert Osazuwa and Larson, Jonathan},
  journal={arXiv preprint arXiv:2404.16130},
  year={2024}
}

@inproceedings{li2025searcho1,
  title={Search-o1: Agentic search-enhanced large reasoning models},
  author={Li, Xiaoxi and Dong, Guanting and Jin, Jiajie and Zhang, Yuyao and Zhou, Yujia and Zhu, Yutao and Zhang, Peitian and Dou, Zhicheng},
  booktitle={Proceedings of the 2025 Conference on Empirical Methods in Natural Language Processing},
  pages={5420--5438},
  year={2025}
}

@inproceedings{cohan2019scicite,
  title={Structural scaffolds for citation intent classification in scientific publications},
  author={Cohan, Arman and Ammar, Waleed and Van Zuylen, Madeleine and Cady, Field},
  booktitle={Proceedings of the 2019 conference of the North American chapter of the Association for Computational Linguistics: human language technologies, volume 1 (long and short papers)},
  pages={3586--3596},
  year={2019}
}

@inproceedings{jantsch2025finecite,
  title={FINECITE: A Novel Approach For Fine-Grained Citation Context Analysis},
  author={Jantsch, Lasse M and Koh, Dong-Jae and Yoon, Seonghwan and Lee, Jisu and Lauscher, Anne and Suh, Young-Kyoon},
  booktitle={Findings of the Association for Computational Linguistics: ACL 2025},
  pages={24525--24542},
  year={2025}
}

@misc{jia2024ddm,
      title={Leveraging Large Language Models for Semantic Query Processing in a Scholarly Knowledge Graph}, 
      author={Runsong Jia and Bowen Zhang and Sergio J. Rodríguez Méndez and Pouya G. Omran},
      year={2026},
      eprint={2405.15374},
      archivePrefix={arXiv},
      primaryClass={cs.IR},
      url={https://arxiv.org/abs/2405.15374}, 
}

@inproceedings{li2024corpuslm,
  title={Corpuslm: Towards a unified language model on corpus for knowledge-intensive tasks},
  author={Li, Xiaoxi and Dou, Zhicheng and Zhou, Yujia and Liu, Fangchao},
  booktitle={Proceedings of the 47th International ACM SIGIR Conference on Research and Development in Information Retrieval},
  pages={26--37},
  year={2024}
}

@article{li2025webthinker,
  title={Webthinker: Empowering large reasoning models with deep research capability},
  author={Li, Xiaoxi and Jin, Jiajie and Dong, Guanting and Qian, Hongjin and Wu, Yongkang and Wen, Ji-Rong and Zhu, Yutao and Dou, Zhicheng},
  journal={arXiv preprint arXiv:2504.21776},
  year={2025}
}

@inproceedings{viswanathan2021citation,
  title={CitationIE: Leveraging the citation graph for scientific information extraction},
  author={Viswanathan, Vijay and Neubig, Graham and Liu, Pengfei},
  booktitle={Proceedings of the 59th Annual Meeting of the Association for Computational Linguistics and the 11th International Joint Conference on Natural Language Processing (Volume 1: Long Papers)},
  pages={719--731},
  year={2021}
}

@article{yao2022react,
  title={React: Synergizing reasoning and acting in language models},
  author={Yao, Shunyu and Zhao, Jeffrey and Yu, Dian and Du, Nan and Shafran, Izhak and Narasimhan, Karthik and Cao, Yuan},
  journal={arXiv preprint arXiv:2210.03629},
  year={2022}
}

@inproceedings{wang2023query2doc,
  title={Query2doc: Query expansion with large language models},
  author={Wang, Liang and Yang, Nan and Wei, Furu},
  booktitle={Proceedings of the 2023 Conference on Empirical Methods in Natural Language Processing},
  pages={9414--9423},
  year={2023}
}

@article{jurgens2018measuring,
  title={Measuring the evolution of a scientific field through citation frames},
  author={Jurgens, David and Kumar, Srijan and Hoover, Raine and McFarland, Dan and Jurafsky, Dan},
  journal={Transactions of the Association for Computational Linguistics},
  volume={6},
  pages={391--406},
  year={2018}
}

@inproceedings{jaradeh2019orkg,
  title={Open research knowledge graph: Next generation infrastructure for semantic scholarly knowledge},
  author={Jaradeh, Mohamad Yaser and Oelen, Allard and Farfar, Kheir Eddine and Prinz, Manuel and D'Souza, Jennifer and Kismih{\'o}k, G{\'a}bor and Stocker, Markus and Auer, S{\"o}ren},
  booktitle={Proceedings of the 10th international conference on knowledge capture},
  pages={243--246},
  year={2019}
}

@article{groth2010nanopublication,
  title={The anatomy of a nanopublication},
  author={Groth, Paul and Gibson, Andrew and Velterop, Jan},
  journal={Information services and use},
  volume={30},
  number={1-2},
  pages={51--56},
  year={2010},
  publisher={SAGE Publications Sage UK: London, England}
}

@article{teufel2002summarizing,
  title={Summarizing scientific articles: experiments with relevance and rhetorical status},
  author={Teufel, Simone and Moens, Marc},
  journal={Computational linguistics},
  volume={28},
  number={4},
  pages={409--445},
  year={2002},
  publisher={MIT Press One Rogers Street, Cambridge, MA 02142-1209, USA journals-info~…}
}

@inproceedings{teufel2009towards,
  title={Towards domain-independent argumentative zoning: Evidence from chemistry and computational linguistics},
  author={Teufel, Simone and Siddharthan, Advaith and Batchelor, Colin},
  booktitle={Proceedings of the 2009 conference on empirical methods in natural language processing},
  pages={1493--1502},
  year={2009}
}

@inproceedings{lauscher2018argument,
  title={An argument-annotated corpus of scientific publications},
  author={Lauscher, Anne and Glava{\v{s}}, Goran and Ponzetto, Simone Paolo},
  booktitle={Proceedings of the 5th Workshop on Argument Mining},
  pages={40--46},
  year={2018}
}

@article{stab2017parsing,
  title={Parsing argumentation structures in persuasive essays},
  author={Stab, Christian and Gurevych, Iryna},
  journal={Computational Linguistics},
  volume={43},
  number={3},
  pages={619--659},
  year={2017}
}

@article{zheng2023judging,
  title={Judging llm-as-a-judge with mt-bench and chatbot arena},
  author={Zheng, Lianmin and Chiang, Wei-Lin and Sheng, Ying and Zhuang, Siyuan and Wu, Zhanghao and Zhuang, Yonghao and Lin, Zi and Li, Zhuohan and Li, Dacheng and Xing, Eric and others},
  journal={Advances in neural information processing systems},
  volume={36},
  pages={46595--46623},
  year={2023}
}

@inproceedings{chiang2023can,
  title={Can large language models be an alternative to human evaluations?},
  author={Chiang, Cheng-Han and Lee, Hung-yi},
  booktitle={Proceedings of the 61st Annual Meeting of the Association for Computational Linguistics (Volume 1: Long Papers)},
  pages={15607--15631},
  year={2023}
}

@article{dubois2023alpacafarm,
  title={Alpacafarm: A simulation framework for methods that learn from human feedback},
  author={Dubois, Yann and Li, Chen Xuechen and Taori, Rohan and Zhang, Tianyi and Gulrajani, Ishaan and Ba, Jimmy and Guestrin, Carlos and Liang, Percy S and Hashimoto, Tatsunori B},
  journal={Advances in Neural Information Processing Systems},
  volume={36},
  pages={30039--30069},
  year={2023}
}

@inproceedings{wang2024large,
  title={Large language models are not fair evaluators},
  author={Wang, Peiyi and Li, Lei and Chen, Liang and Cai, Zefan and Zhu, Dawei and Lin, Binghuai and Cao, Yunbo and Kong, Lingpeng and Liu, Qi and Liu, Tianyu and others},
  booktitle={Proceedings of the 62nd Annual Meeting of the Association for Computational Linguistics (Volume 1: Long Papers)},
  pages={9440--9450},
  year={2024}
}

@article{panickssery2024llm,
  title={Llm evaluators recognize and favor their own generations},
  author={Panickssery, Arjun and Bowman, Samuel R and Feng, Shi},
  journal={Advances in Neural Information Processing Systems},
  volume={37},
  pages={68772--68802},
  year={2024}
}

@inproceedings{carbonell1998use,
  title={The use of MMR, diversity-based reranking for reordering documents and producing summaries},
  author={Carbonell, Jaime and Goldstein, Jade},
  booktitle={Proceedings of the 21st annual international ACM SIGIR conference on Research and development in information retrieval},
  pages={335--336},
  year={1998}
}

@inproceedings{athar2011sentiment,
  title={Sentiment analysis of citations using sentence structure-based features},
  author={Athar, Awais},
  booktitle={Proceedings of the ACL 2011 student session},
  pages={81--87},
  year={2011}
}

@article{taffa2023leveraging,
  title={Leveraging llms in scholarly knowledge graph question answering},
  author={Taffa, Tilahun Abedissa and Usbeck, Ricardo},
  journal={arXiv preprint arXiv:2311.09841},
  year={2023}
}

@article{bezerra2025leveraging,
  title={Leveraging GANs for citation intent classification and its impact on citation network analysis},
  author={Bezerra, Davi A and Silva, Filipi N and Amancio, Diego R},
  journal={arXiv preprint arXiv:2505.21162},
  year={2025}
}

@article{ma2025citation,
  title={Citation recommendation based on argumentative zoning of user queries},
  author={Ma, Shutian and Zhang, Chengzhi and Zhang, Heng and Gao, Zheng},
  journal={Journal of Informetrics},
  volume={19},
  number={1},
  pages={101607},
  year={2025},
  publisher={Elsevier}
}

@inproceedings{schrader2023mulms,
  title={MuLMS-AZ: an argumentative zoning dataset for the materials science domain},
  author={Schrader, Timo Pierre and B{\"u}rkle, Teresa and Henning, Sophie and Tan, Sherry and Finco, Matteo and Gr{\"u}newald, Stefan and Indrikova, Maira and Hildebrand, Felix and Friedrich, Annemarie},
  booktitle={Proceedings of the 4th Workshop on Computational Approaches to Discourse (CODI 2023)},
  pages={1--15},
  year={2023}
}

@article{landis1977measurement,
  title={The measurement of observer agreement for categorical data},
  author={Landis, J Richard and Koch, Gary G},
  journal={biometrics},
  pages={159--174},
  year={1977},
  publisher={JSTOR}
}

@inproceedings{lahiri2023citeprompt,
  title={Citeprompt: Using prompts to identify citation intent in scientific papers},
  author={Lahiri, Avishek and Sanyal, Debarshi Kumar and Mukherjee, Imon},
  booktitle={2023 ACM/IEEE Joint Conference on Digital Libraries (JCDL)},
  pages={51--55},
  year={2023},
  organization={IEEE}
}

@article{paolini2025citefusion,
  title={CiteFusion: an ensemble framework for citation intent classification harnessing dual-model binary couples and SHAP analyses},
  author={Paolini, Lorenzo and Vahdati, Sahar and Di Iorio, Angelo and Wardenga, Robert and Heibi, Ivan and Peroni, Silvio},
  journal={Scientometrics},
  pages={1--71},
  year={2025},
  publisher={Springer}
}

@inproceedings{shui2024fine,
  title={Fine-tuning language models on multiple datasets for citation intention classification},
  author={Shui, Zeren and Karypis, Petros and Karls, Daniel S and Wen, Mingjian and Manchanda, Saurav and Tadmor, Ellad B and Karypis, George},
  booktitle={Findings of the association for computational linguistics: EMNLP 2024},
  pages={16718--16732},
  year={2024}
}

\appendix

\section{Taxonomy Mapping to Prior Schemes}
\label{app:taxonomy_mapping}

Our four-class taxonomy (\textsc{Critique}, \textsc{Adoption}, \textsc{Benchmark}, \textsc{Neutral}) is derived from the citation-intent literature surveyed in \S\ref{sec:rw_intent} and adapted to the unit at which the downstream tasks in \S\ref{sec:tasks} operate. This appendix specifies the correspondence to two reference schemes --- ACL-ARC~\citep{jurgens2018measuring} and SciCite~\citep{cohan2019scicite} --- and explains the conceptual moves that distinguish our taxonomy from theirs.

\begin{table}[htbp]
\centering
\small
\setlength{\tabcolsep}{3.5pt} 
\begin{tabularx}{\columnwidth}{@{} l >{\raggedright\arraybackslash}X >{\raggedright\arraybackslash}X @{}}
\toprule
\textbf{Our class} & \textbf{ACL-ARC} & \textbf{SciCite} \\
\midrule
\textsc{Critique} & \textit{CompareOrContrast} (negative); \textit{Motivation} (gap-identifying) & \textit{Result Comparison} (negative) \\
\addlinespace
\textsc{Adoption} & \textit{Uses}; \textit{Extends} & \textit{Method} \\
\addlinespace
\textsc{Benchmark} & \textit{CompareOrContrast} (comparative) & \textit{Result Comparison} (neutral) \\
\addlinespace
\textsc{Neutral} & \textit{Background}; \textit{Future} & \textit{Background} \\
\bottomrule
\end{tabularx}
\caption{Correspondence between the four-class claim taxonomy and the ACL-ARC and SciCite citation-intent schemes. Italicised qualifiers identify the subcase of the source class that maps to our class when the source class is split across two of ours.}
\label{tab:taxonomy_mapping}
\end{table}

\paragraph{Adoption.} The most direct mapping. Both reference schemes contain a class for citations in which the source paper takes a method, dataset, or artefact from the cited work as part of its own contribution. ACL-ARC's \textit{Uses} and \textit{Extends} together cover this; SciCite folds them into \textit{Method}. Our \textsc{Adoption} corresponds to this region of intent space with no further refinement.

\paragraph{Neutral.} Maps onto ACL-ARC's \textit{Background} and \textit{Future} together, and onto the corresponding region of SciCite's \textit{Background}. These citations are non-evaluative: the source paper mentions the target for contextual, historical, or future-work reasons without taking a stance on its quality, utility, or comparability.

\paragraph{Critique and Benchmark.} Both classes are derived from a single class in each reference scheme. ACL-ARC groups all forms of side-by-side discussion under \textit{CompareOrContrast}; SciCite groups numerical comparison under \textit{Result Comparison}. Our taxonomy splits each along the dimension of evaluative valence: a comparison that highlights limitations of the cited work, or that motivates the source paper's contribution by identifying a gap, is a \textsc{Critique}; a comparison that places the cited work alongside others as a quantitative or methodological reference point is a \textsc{Benchmark}.

\paragraph{Methodological moves.} Two design choices distinguish our taxonomy from the reference schemes. The first is the split just described. This split is motivated by the downstream tasks the network supports. Consensus report generation (\S\ref{sec:task2}) aggregates incoming claims by type into a structured \textit{Strengths / Limitations / Use as Baseline / Reception} rollup; the rollup is not well-defined under a single \textit{CompareOrContrast} or \textit{Result Comparison} class because the same class would feed into multiple rollups in different ways. Separating the evaluative-valence dimension at extraction time makes the aggregation operation well-defined. Graph analytics (\S\ref{sec:task3}) shows the same dependency: the \textit{Most Critiqued} and \textit{Benchmark Status} queries operate on disjoint subgraphs and would collapse into a single ambiguous ranking under either reference scheme.

The second is the renaming of \textit{Result Comparison} to \textsc{Benchmark}. The renamed class is broader than literal numerical comparison: a source paper can treat a target as a benchmark when discussing it as a comparative reference point outside a results table, and \textit{benchmark} is the term used in the empirical literatures from which our scholarly instantiation draws. The rename also aligns the class name with the corresponding query family in \S\ref{sec:task3} (\textit{Benchmark Status}), where the unit being ranked is the role of ``paper treated as comparative reference''.

\paragraph{Interpretive caveat.} The mapping above is conceptual rather than mechanical. Individual citations in the ACL-ARC and SciCite test sets exhibit ambiguity along the same boundaries our taxonomy resolves: a \textit{CompareOrContrast} citation in ACL-ARC may be a \textsc{Critique} or a \textsc{Benchmark} in our scheme depending on whether the comparison is evaluative. We do not claim that a classifier trained on ACL-ARC or SciCite labels would transfer to our taxonomy by relabelling alone; the unit of classification (paragraph-level aggregated context rather than sentence) and the evaluative-valence dimension would both need to be handled explicitly.

The attitude axis (\S\ref{sec:extraction}) has no direct correspondent in ACL-ARC or SciCite, both of which omit citation polarity from their schemes; the closest prior reference for the polarity dimension is the citation-sentiment literature initiated by~\citet{athar2011sentiment}.

\section{Extraction Pipeline Implementation Details}
\label{app:implementation}

This appendix documents the prompt, the structured-output schema, and the
SPARQL queries used to build the claim network of \S\ref{sec:claim_network}.
Together they constitute the reproducibility floor for the extraction
pipeline: any implementation that issues the same prompt against a model
honouring the same schema, over a triplestore queried with the same SPARQL,
will reproduce the claim set up to the non-determinism of the language
model itself. The pipeline is implemented in Python with asynchronous
\texttt{httpx} against the DashScope OpenAI-compatible endpoint;
released code accompanies the paper.

\subsection{Extraction Prompt}
\label{app:b1_prompt}

The extraction is a single LLM call per ordered paper pair $(p_s, p_t)$.
The call uses Qwen3-Max via DashScope with deterministic decoding
(\texttt{temperature=0.0}) and JSON-mode structured output
(\texttt{response\_format=\{"type":\,"json\_object"\}}). The prompt
consists of a fixed system turn and a per-pair user turn. No external
few-shot turns are appended; in-prompt demonstrations are carried inside
the two turns themselves.

\paragraph{System prompt.} The system prompt fixes the output shape and
the multi-label / single-label distinction between \texttt{claim\_types}
and \texttt{attitude}, illustrated by a two-claim JSON exemplar:

\begin{quote}\small
\begin{lstlisting}[
  breaklines=true,
  breakatwhitespace=true,
  basicstyle=\ttfamily\small,
  columns=flexible,
  keepspaces=true,
  frame=none,
  xleftmargin=0pt,
  xrightmargin=0pt,
  aboveskip=0pt,
  belowskip=0pt
]
You are an expert academic data extractor performing
MULTI-LABEL classification.
Output STRICTLY in JSON format.
Your response MUST be a valid JSON object containing
a SINGLE key named "claims", which is a list of objects.
Each claim MUST have:
- a `claim_types` field - a JSON ARRAY of one or more
  labels (NEVER a bare string).
- an `attitude` field - a JSON STRING that is EXACTLY
  one of: "Positive", "Negative", "Neutral"
  (single-label, never an array).
Example format:
{
  "claims":[
    {
      "window_index": 1,
      "specific_claim": "We adopt PointNet++'s set abstraction layer with radius 0.2 and report 92.4% mIoU vs PointNet++'s 90.7% on ScanNet.",
      "claim_types": ["Adoption", "Benchmark"],
      "attitude": "Positive"
    },
    {
      "window_index": 2,
      "specific_claim": "PointNet++ suffers from O(N^2) neighborhood query cost on large scenes.",
      "claim_types": ["Critique"],
      "attitude": "Negative"
    }
  ]
}
\end{lstlisting}
\end{quote}

\paragraph{User-turn template.} The user turn is formed by substituting
three placeholders: \texttt{\{source\_title\}} (the title of the citing
paper $p_s$); \texttt{\{target\_text\}} (the bibliographic text of the
cited reference $p_t$, truncated to 250 characters at SPARQL retrieval
time as described in Appendix~\ref{app:b3_sparql}); and
\texttt{\{combined\_context\}}, the deduplicated citing paragraphs from
$p_s$ that invoke $p_t$, joined in the form
``\texttt{[Window 1]: \ldots\textbackslash n\textbackslash n[Window 2]: \ldots}''.

\begin{quote}\small
\begin{lstlisting}[
  breaklines=true,
  breakatwhitespace=true,
  basicstyle=\ttfamily\small,
  columns=flexible,
  keepspaces=true,
  frame=none,
  xleftmargin=0pt,
  xrightmargin=0pt,
  aboveskip=0pt,
  belowskip=0pt
]
I will provide you with distinct citation contexts where
[Source] cites [Target].
For EACH window independently, extract ONE highly specific
claim, assign it a MULTI-LABEL set of `claim_types`, and a
SINGLE-LABEL `attitude`.

[Source]: {source_title}
[Target Reference]: {target_text}

=== CONTEXT WINDOWS ===
{combined_context}

=== EXTRACTION RULES ===
1. Be extremely specific. Preserve exact flaws,
   hyperparameter values, dataset names, metric numbers
   (e.g. "mIoU 73.4% on S3DIS Area-5",
   "voxel size 0.05 m",
   "8-layer MLP with hidden dim 256").
2. Stay grounded in the window text - do NOT invent
   numbers or methods that are not stated.
3. If the window does not actually discuss the [Target]
   (e.g. it cites the target alongside many others
   without saying anything about it), produce a short
   claim describing that situation and label it
   ["Neutral"].

=== LABEL DEFINITIONS (multi-label allowed) ===
- Critique: Highlights a flaw, limitation, computational
  cost, accuracy ceiling, or missing capability in the
  Target.
- Adoption: Source uses, inherits, re-implements, or
  builds upon a method, module, architecture, training
  recipe, or hyperparameter from the Target.
- Benchmark: Source quantitatively compares its own
  metrics (Accuracy, mIoU, F1, latency, params, FLOPs,
  etc.) against the Target's reported numbers, or uses
  the Target as a baseline in a results table.
- Neutral: Casual co-mention without attitude or detail;
  OR target listed in a generic background list. Use
  ONLY when none of the above three apply.

=== MULTI-LABEL GUIDANCE ===
A single claim CAN and SHOULD carry multiple labels when
supported:
- "We follow KPConv's deformable kernel and outperform
   it by 1.8 mIoU"  -> ["Adoption", "Benchmark"]
- "Although Transformer-based, PointTransformer's
   quadratic attention limits scalability; we adopt its
   self-attention block but replace softmax with linear
   attention"  -> ["Critique", "Adoption"]
- "PointNet's max-pool aggregation discards local
   geometry; our method recovers 4.2 mIoU over PointNet
   on ScanNet"  -> ["Critique", "Benchmark"]
- "We adopt the same training schedule (200 epochs,
   cosine LR)"  -> ["Adoption"]
- "Neutral" is mutually exclusive - never combine it
  with other labels.

=== ATTITUDE DEFINITIONS (single-label, EXACTLY one) ===
- Positive: Source endorses, praises, builds upon
  approvingly, or reports favorable comparison toward
  the Target (e.g. "we follow the strong baseline X",
  "outperform X by adopting its core idea Y",
  "X's elegant design inspires our module").
- Negative: Source criticizes, points out
  flaws/limitations, or reports the Target as inferior
  (e.g. "X suffers from quadratic cost", "X fails on
  large scenes", "we surpass X by 4 mIoU due to X's
  max-pool bottleneck").
- Neutral: No clear stance - pure factual citation,
  background mention, or balanced description without
  endorsement or criticism.

=== ATTITUDE vs CLAIM_TYPES ===
- attitude and claim_types are INDEPENDENT axes.
  A "Benchmark" claim can be Positive (we beat X) or
  Negative (X still ahead) or Neutral (head-to-head
  report w/o judgment).
- "Adoption" alone usually maps to Positive (reuse
  implies endorsement) unless the text frames the reuse
  critically.
- "Critique" alone almost always Negative.
- claim_types == ["Neutral"] usually pairs with
  attitude "Neutral".

Output the JSON object now.
\end{lstlisting}
\end{quote}

\paragraph{In-prompt demonstrations.} Four inline exemplars appear under
the \textsc{Multi-Label Guidance} block, covering the four label-set
combinations that recur across the corpus (\textsc{Adoption} +
\textsc{Benchmark}; \textsc{Critique} + \textsc{Adoption};
\textsc{Critique} + \textsc{Benchmark}; \textsc{Adoption} alone). One
further two-claim JSON exemplar is embedded in the system prompt above,
demonstrating the wire shape of the output. The model is never given
out-of-corpus examples; demonstrations stay close to the 3D point cloud
literature so that surface-level style does not drift from the target
domain. The prompt reproduced here is the final version, produced after
one iteration that lifted the schema from a single \texttt{claim\_type}
label to the multi-label \texttt{claim\_types} array described in
Appendix~\ref{app:b2_schema}; all claims reported in
\S\ref{sec:stats} were extracted under this version.

\subsection{Structured Output Schema}
\label{app:b2_schema}

The model's response is validated against a Pydantic schema before it
is accepted into the network. The schema fixes both the field shapes
and the closed enumerations of allowed values, so the extraction is
type-safe end-to-end.

\paragraph{Schema definition.}

\begin{quote}\small
\begin{lstlisting}[
  breaklines=true,
  breakatwhitespace=true,
  basicstyle=\ttfamily\small,
  columns=flexible,
  keepspaces=true,
  frame=none,
  xleftmargin=0pt,
  xrightmargin=0pt,
  aboveskip=0pt,
  belowskip=0pt
]
from pydantic import BaseModel, Field
from typing import List, Literal

ClaimLabel = Literal[
    "Critique", "Adoption", "Benchmark", "Neutral"
]
ClaimAttitude = Literal[
    "Positive", "Negative", "Neutral"
]


class ClassifiedClaim(BaseModel):
    window_index: int = Field(
        ...,
        description="The index of the context window "
                    "(1-based)."
    )
    specific_claim: str = Field(
        ...,
        description="The highly specific extracted "
                    "claim, usage, or critique."
    )
    claim_types: List[ClaimLabel] = Field(
        ...,
        min_length=1,
        description=(
            "Multi-label set. A claim can have multiple "
            "labels simultaneously (e.g. "
            "['Adoption', 'Benchmark'] when the source "
            "both reuses a method AND compares metrics "
            "against the target). Use ['Neutral'] only "
            "when no other label fits."
        ),
    )
    attitude: ClaimAttitude = Field(
        ...,
        description=(
            "Single-label attitude of the Source toward "
            "the Target as expressed in this window. "
            "EXACTLY one of: 'Positive', 'Negative', "
            "'Neutral'."
        ),
    )


class ClaimExtraction(BaseModel):
    claims: List[ClassifiedClaim]
\end{lstlisting}
\end{quote}

\paragraph{Field summary.} Table~\ref{tab:b2_fields} summarises the
fields, their cardinalities, and their allowed values.

\begin{table}[h]
\centering
\small
\begin{tabular}{p{1.9cm}p{1.4cm}p{3.0cm}}
\toprule
\textbf{Field} & \textbf{Type} & \textbf{Allowed values / notes} \\
\midrule
\texttt{claims} & list & top-level wrapper; may be empty \\
\addlinespace
\texttt{window\_index} & int & 1-based; aligns each claim with its source paragraph \\
\addlinespace
\texttt{specific\_claim} & string & non-empty free text \\
\addlinespace
\texttt{claim\_types} & list of enum & one or more of \{Critique, Adoption, Benchmark, Neutral\}; Neutral mutually exclusive by prompt instruction \\
\addlinespace
\texttt{attitude} & enum & exactly one of \{Positive, Negative, Neutral\}; independent of \texttt{claim\_types} \\
\bottomrule
\end{tabular}
\caption{Output schema fields. The Neutral mutual-exclusion is enforced by the prompt rather than by the schema; downstream injection applies a defensive filter that drops claims whose label set is empty after intersection with the four-class enum.}
\label{tab:b2_fields}
\end{table}

\paragraph{Validation failure behaviour.} The raw response is parsed
with \texttt{ClaimExtraction.model\_validate\_json(\ldots)}. If parsing
raises for any reason --- malformed JSON, missing key, type mismatch,
value outside the \texttt{Literal} enumeration, or empty
\texttt{claim\_types} --- the call enters a retry loop with
\texttt{MAX\_RETRIES = 3} attempts per pair, separated by a two-second
\texttt{asyncio.sleep} and using the identical prompt. Because decoding
is deterministic the retry is in effect a network-level resilience
mechanism rather than a sampling mechanism. Markdown code fences are
stripped before validation in case the model ignores the JSON-mode
constraint and returns its output inside triple backticks. Quota- or
billing-related errors (substring match on
\texttt{"quota"} / \texttt{"balance"} / \texttt{"arrears"}) abort the
run immediately rather than entering the retry loop, so that an
exhausted credit balance is surfaced rather than absorbed silently.
After three failures the pair is logged and dropped from the output;
no unvalidated claims are written. A second defensive filter at
injection time removes any claim whose \texttt{claim\_types} resolves
to an empty set after intersection with the four-class enum and coerces
any unrecognised \texttt{attitude} value to \texttt{Neutral}. The
Pydantic enums should already exclude both cases; the injection-time
filter is a belt-and-braces check.

\subsection{SPARQL Queries Used in Extraction}
\label{app:b3_sparql}

The extraction-time queries operate against the GraphDB triplestore
holding the DDM representation of the corpus (\S\ref{sec:ddm}). All
queries share the following prefixes:

\begin{quote}\small
\begin{lstlisting}[
  breaklines=true,
  breakatwhitespace=true,
  basicstyle=\ttfamily\small,
  columns=flexible,
  keepspaces=true,
  frame=none,
  xleftmargin=0pt,
  xrightmargin=0pt,
  aboveskip=0pt,
  belowskip=0pt
]
PREFIX askg-onto:
  <https://www.anu.edu.au/onto/scholarly#>
PREFIX domo:
  <http://example.org/domo/>
PREFIX dc:
  <http://purl.org/dc/elements/1.1/>
\end{lstlisting}
\end{quote}

\paragraph{Paper $\to$ Section $\to$ Paragraph $\to$ Citation
traversal.} A single \texttt{SELECT} query walks all four hops of the
DDM hierarchy, applies the bibliography filter (see below) and the
citation-marker resolver (see below), and returns one row per
(source paper, target reference, citing paragraph) triple. Rows are
then grouped in Python on $(p_s, p_t)$ to form the per-pair window
list.

\begin{quote}\small
\begin{lstlisting}[
  breaklines=true,
  breakatwhitespace=true,
  basicstyle=\ttfamily\small,
  columns=flexible,
  keepspaces=true,
  frame=none,
  xleftmargin=0pt,
  xrightmargin=0pt,
  aboveskip=0pt,
  belowskip=0pt
]
SELECT ?sourceUri ?sourceTitle
       ?targetUri ?targetText ?paraText
WHERE {
    # 1. Source paper + its title
    ?sourceUri a askg-onto:Paper ;
               dc:title ?sourceTitle ;
               askg-onto:hasSection ?sec .

    # 2. Exclude reference / bibliography sections
    OPTIONAL { ?sec dc:title ?secTitle }
    FILTER (!BOUND(?secTitle) ||
            !REGEX(STR(?secTitle),
                   "reference|bibliography", "i"))

    # 3. Paragraph and its raw citation marker
    #    (may be IRI or numeric string)
    ?sec  askg-onto:hasParagraph ?para .
    ?para askg-onto:hasCitation ?rawCitation ;
          domo:Text ?paraText .

    # 4. Resolve raw marker -> canonical Reference IRI
    BIND(
      IF(isIRI(?rawCitation),
         ?rawCitation,
         IRI(CONCAT(STR(?sourceUri),
                    "-Reference-",
                    REPLACE(STR(?rawCitation),
                            "[^0-9]", "")))
      ) AS ?targetUri
    )

    # 5. Target reference text on the resolved node
    ?targetUri domo:Text ?targetText .

    # 6. Drop trivially short paragraphs
    FILTER(STRLEN(STR(?paraText)) > 20)
}
\end{lstlisting}
\end{quote}

After the query returns, the Python driver groups rows by $(p_s, p_t)$, truncates \texttt{?targetText} to 250 characters for the prompt, and deduplicates windows. Two additional regex filters strip residual PDF-extraction noise before any window is sent to the LLM: paragraphs containing the substring \texttt{<span id="page-} (\texttt{marker}'s inline page-boundary spans that survived PDF $\to$ Markdown conversion) and paragraphs matching \texttt{\^{}\textbackslash s*[-*]?\textbackslash s*(<[\^{}>]+>)* \textbackslash s*\textbackslash[\textbackslash d+\textbackslash] \textbackslash s+[A-Z]} (bibliography entries that escaped the SPARQL filter because they were misclassified as body paragraphs at ingestion time).

\paragraph{Bibliography exclusion.} Bibliography exclusion is not a
separate query; it is the \texttt{FILTER} on \texttt{?secTitle} embedded
in the traversal above. The clause

\begin{quote}\small
\begin{lstlisting}[
  breaklines=true,
  breakatwhitespace=true,
  basicstyle=\ttfamily\small,
  columns=flexible,
  keepspaces=true,
  frame=none,
  xleftmargin=0pt,
  xrightmargin=0pt,
  aboveskip=0pt,
  belowskip=0pt
]
OPTIONAL { ?sec dc:title ?secTitle }
FILTER (!BOUND(?secTitle) ||
        !REGEX(STR(?secTitle),
               "reference|bibliography", "i"))
\end{lstlisting}
\end{quote}

drops any \texttt{Section} whose title case-insensitively contains
\texttt{reference} or \texttt{bibliography}, while keeping sections that
carry no \texttt{dc:title} at all. The \texttt{!BOUND} disjunct is
deliberate: at DDM ingestion time a small number of section nodes are
created without a recoverable heading, and excluding all
title-less sections wholesale would lose body content. The
line-level regex filter in the Python driver provides a second layer
of defence against bibliography entries that have been misclassified
upstream.

\paragraph{Citation-marker resolution.} The DDM corpus stores
\texttt{askg-onto:hasCitation} heterogeneously: some papers carry an
IRI pointing directly at the target \texttt{Reference} node, while
others carry a literal string such as \texttt{"10"} or
\texttt{"[10]"}. The inline \texttt{BIND} above branches on
\texttt{isIRI(?rawCitation)}: an IRI is passed through unchanged,
while a literal is converted into the canonical reference IRI by
concatenating the source paper IRI with \texttt{"-Reference-"} and
the digits of the marker. The subsequent triple
\texttt{?targetUri domo:Text ?targetText} succeeds only when the
reconstructed IRI matches an actual \texttt{Reference} node in the
triplestore, which doubles as a sanity check on the resolver: an
unresolvable marker produces no row rather than a row with a phantom
target.

\section{Task 1 Sensitivity Sweep: Full Per-Cell Numbers}
\label{app:sensitivity}

Table~\ref{tab:sensitivity_full} reports the complete per-cell numbers underlying Table~\ref{tab:task1_sensitivity}: win rate, delta from default, and 4-trial Agree\% (read as mean signed weighted margin on the \textit{Overall (weighted)} column).

\begin{table*}[h]
\centering
\scriptsize
\setlength{\tabcolsep}{2.5pt}
\begin{tabular}{l l | c c | c c | c c | c c | c c}
\toprule
& & \multicolumn{2}{c|}{Relev} & \multicolumn{2}{c|}{Factual} & \multicolumn{2}{c|}{MSI} & \multicolumn{2}{c|}{Conc} & \multicolumn{2}{c}{Overall (wtd)} \\
\textbf{Axis} & \textbf{Val} & WR & Agr & WR & Agr & WR & Agr & WR & Agr & WR & Margin \\
\midrule
$\tau_\text{floor}$  & 0.15 & 0.538 & 0.133 & 0.667 & 0.000 & 0.545 & 0.267 & 0.538 & 0.533 & 0.533 & 0.092 \\
$\tau_\text{floor}$  & \textbf{0.25} & 0.333 & 0.267 & 1.000 & 0.733 & 0.714 & 0.467 & 0.143 & 0.467 & 0.700 & 0.044 \\
$\tau_\text{floor}$  & 0.35 & 0.667 & 0.267 & 0.200 & 0.000 & 0.800 & 0.267 & 0.357 & 0.533 & 0.667 & 0.098 \\
\midrule
$\tau_\text{anchor}$ & 0.20 & 0.615 & 0.133 & 0.800 & 0.000 & 0.692 & 0.200 & 0.273 & 0.400 & 0.667 & 0.208 \\
$\tau_\text{anchor}$ & \textbf{0.30} & 0.333 & 0.267 & 1.000 & 0.733 & 0.714 & 0.467 & 0.143 & 0.467 & 0.700 & 0.044 \\
$\tau_\text{anchor}$ & 0.40 & 0.583 & 0.333 & 0.400 & 0.000 & 0.556 & 0.400 & 0.385 & 0.467 & 0.533 & 0.020 \\
\midrule
$w_\text{claim}$     & 0.20 & 0.571 & 0.200 & 0.375 & 0.000 & 0.444 & 0.133 & 0.667 & 0.400 & 0.600 & -0.020 \\
$w_\text{claim}$     & \textbf{0.40} & 0.333 & 0.267 & 1.000 & 0.733 & 0.714 & 0.467 & 0.143 & 0.467 & 0.700 & 0.044 \\
$w_\text{claim}$     & 0.60 & 0.714 & 0.133 & 0.250 & 0.000 & 0.727 & 0.333 & 0.545 & 0.267 & 0.786 & 0.161 \\
\bottomrule
\end{tabular}
\caption{Full per-cell sensitivity-sweep numbers. \textbf{WR} is win rate over non-tied items; \textbf{Agr} is 4-trial Agree\% (the fraction of items on which all four trials returned the same non-tie verdict); \textbf{Margin} on the \textit{Overall (weighted)} column is the mean signed weighted margin per item, with range $[-1, +1]$, following the convention established in \S\ref{sec:eval_protocol}. The most diagnostic pattern is that the \textit{Factual} Agree\% collapses to 0.000 on every perturbation, indicating that the large \textit{Factual Correctness} win-rate swings under the sweep correspond to operating points at which no item achieves a unanimous 4-trial verdict on that dimension.}
\label{tab:sensitivity_full}
\end{table*}

\section{Task 2 MMR Cap Ablation: Per-Target Numbers}
\label{app:mmr_ablation}

Table~\ref{tab:mmr_full} reports the per-target absolute rubric scores and prompt-size statistics underlying Table~\ref{tab:task2_mmr_ablation}.

\begin{table*}[h]
\centering
\scriptsize
\setlength{\tabcolsep}{3pt}
\begin{tabular}{l l | c c c c c | c c c c c | c c c c c}
\toprule
& & \multicolumn{5}{c|}{$M = 5$} & \multicolumn{5}{c|}{$M = 15$ (default)} & \multicolumn{5}{c}{$M = \text{all}$} \\
\textbf{ID} & \textbf{Target} & cl & src & ID & SD & PC & cl & src & ID & SD & PC & cl & src & ID & SD & PC \\
\midrule
C1  & Point Transformer                     & 15 & 7 & 14 & 13 & 13 & 34 &  9 & 13 & 13 & 13 & 37 &  9 & 14 & 13 & 13 \\
C2  & PointNet                              & 15 & 7 & 12 & 13 & 12 & 40 & 16 & 14 & 13 & 14 & 95 & 24 & 14 & 13 & 14 \\
C3  & PointNet++                            & 15 & 8 & 14 & 13 & 13 & 42 & 13 & 14 & 13 & 14 & 72 & 18 & 14 & 13 & 14 \\
C4  & 4D Spatio-Temporal ConvNets           & 15 & 7 & 14 & 13 & 13 & 35 &  9 & 14 & 13 & 14 & 53 &  9 & 14 & 13 & 13 \\
C5  & Dynamic Graph CNN                     & 14 & 8 & 13 & 13 & 13 & 26 & 11 & 14 & 13 & 13 & 34 & 12 & 14 & 13 & 13 \\
C6  & ScanNet                               & 12 & 8 & 14 & 13 & 13 & 24 & 11 & 14 & 13 & 12 & 30 & 11 & 14 & 13 & 13 \\
C7  & ShapeNet                              & 12 & 9 & 14 & 13 & 12 & 27 & 12 & 13 & 13 & 13 & 33 & 12 & 14 & 13 & 12 \\
C8  & PointCNN                              &  3 & 2 & 13 &  7 & 10 &  3 &  2 & 13 & 10 &  9 &  3 &  2 & 13 &  7 &  9 \\
C9  & KPConv                                & 15 & 5 & 12 & 13 & 12 & 27 &  8 & 14 & 13 & 14 & 34 &  8 & 14 & 13 & 13 \\
C10 & Attention Is All You Need             & 13 & 5 & 13 & 13 & 10 & 23 &  7 & 12 & 10 & 12 & 23 &  7 & 12 & 13 & 10 \\
\bottomrule
\end{tabular}
\caption{Per-target MMR cap ablation. \textbf{cl} is the number of claims in the prompt; \textbf{src} is the number of distinct source papers represented; \textbf{ID}/\textbf{SD}/\textbf{PC} are absolute LLM-judge rubric scores for \textit{Information Density}, \textit{Source Diversity}, and \textit{Perspective Coverage}. The $M = 30$ column is omitted from this table for layout reasons but is included in the summary statistics of Table~\ref{tab:task2_mmr_ablation}. C8 (PointCNN) has only three incoming claims in the corpus and is therefore insensitive to the cap by construction; its \textit{Source Diversity} score nevertheless varies between $7$ and $10$ across runs on identical input, illustrating the single-trial judge variance discussed in \S\ref{sec:task2_mmr_ablation}.}
\label{tab:mmr_full}
\end{table*}

\section{Task 3 Query Family Pseudocode}
\label{app:task3_pseudocode}

The pseudocode below documents the operation actually executed for each query family in \S\ref{sec:task3} and \S\ref{sec:experiment_task3}. Notation follows \S\ref{sec:task3}: $G = (V, E)$ is the paper-level multigraph, $n^\tau_{(s,t)}$ is the claim count of type $\tau \in \{\textsc{Cr}, \textsc{Ad}, \textsc{Be}, \textsc{Ne}\}$ on edge $(s, t)$, and $\mathcal{T}(\text{topic}) \subseteq V$ is a hand-curated topic node set.

\paragraph{Influence.}
\begin{quote}\small
\begin{lstlisting}[basicstyle=\ttfamily\small, breaklines=true,
  frame=none, xleftmargin=0pt, columns=flexible]
G_Ad <- subgraph of G keeping edges with n_Ad > 0,
        each weighted by w_e = n_Ad(s,t)
pr   <- weighted_pagerank(G_Ad, damping=0.85)
return top_k(pr restricted to T(topic))
\end{lstlisting}
\end{quote}

\paragraph{Most Critiqued / Benchmark Status / Most Endorsed / Most Rejected.}
\begin{quote}\small
\begin{lstlisting}[basicstyle=\ttfamily\small, breaklines=true,
  frame=none, xleftmargin=0pt, columns=flexible]
for each node t in T(topic):
    score[t] <- sum over incoming edges (s,t):
                  n_tau(s,t)   for the relevant tau
                  (Cr / Be / pos-attitude / neg-attitude)
return top_k(score)
\end{lstlisting}
\end{quote}

\paragraph{Controversy (Polarity).}
\begin{quote}\small
\begin{lstlisting}[basicstyle=\ttfamily\small, breaklines=true,
  frame=none, xleftmargin=0pt, columns=flexible]
for each node t with n_total(t) >= theta (theta = 5):
    rho[t] <- (n_Ad(t) - n_Cr(t)) / n_total(t)
loved        <- top_k(rho ascending = False)
controversial<- top_k(rho ascending = True)
return loved, controversial
\end{lstlisting}
\end{quote}

\paragraph{Bridge.}
\begin{quote}\small
\begin{lstlisting}[basicstyle=\ttfamily\small, breaklines=true,
  frame=none, xleftmargin=0pt, columns=flexible]
for each topic pair (T1, T2):
    for each node v in V:
        a <- |neighbours(v) in T1|   # in- + out-, multigraph
        b <- |neighbours(v) in T2|
        bridge[v] <- min(a, b)
        tie_breaker[v] <- a + b
    return top_k(bridge desc, then tie_breaker desc)
\end{lstlisting}
\end{quote}

\paragraph{Topic Citation Behaviour.}
\begin{quote}\small
\begin{lstlisting}[basicstyle=\ttfamily\small, breaklines=true,
  frame=none, xleftmargin=0pt, columns=flexible]
for each topic T:
    for tau in {Cr, Ad, Be, Ne}:
        total_tau[T] <- sum n_tau(s,t)
                        over edges (s,t) with s in T
    report percentage split across tau, paper count |T|,
    and total outgoing claim count for T
\end{lstlisting}
\end{quote}

\section{Question Set for Task 1}
\label{app:task1_questions}

The fifteen questions evaluated in \S\ref{sec:task1} and reported in Table~\ref{tab:task1} are listed below. The set was hand-curated to exercise the kinds of question that scientific question answering is expected to handle, balanced across five categories rather than skewed toward any one type. The categories are: \textit{limitation} (questions whose answer is the documented weaknesses of a target work, distributed across the citing prose of many later papers); \textit{evolution} (questions whose answer is the lineage of a method, spanning a sequence of adopting and extending papers); \textit{comparison} (questions whose answer is the relative positioning of two or more paradigms within the field); \textit{challenge} (questions whose answer is the community's
collective response to a recurring problem); and \textit{controversy} (questions whose answer requires aggregating mixed and partially opposing stances toward a single work or trend).

The category breakdown is three questions per category, with each question additionally specifying its \textit{target nodes}: the papers in the corpus that the question is principally about. Target nodes are not visible to either system at retrieval time and are used only to anchor the question on a specific region of the corpus during set construction.

\paragraph{Limitation questions.}

\begin{enumerate}[label=Q\arabic*, leftmargin=*, itemsep=0.4em]
\setcounter{enumi}{0}
\item What are the main limitations of PointNet that subsequent research has identified, and how have these limitations influenced the design of later architectures?\\
\textit{Target:} PointNet.
\item What specific technical shortcomings have been raised against sparse convolution approaches like Minkowski Convolutional Neural Networks in the literature?\\
\textit{Target:} MinkowskiNet.
\item Why has KPConv received criticism in the point cloud community despite its novel kernel point design, and what recurring concerns appear across different papers?\\
\textit{Target:} KPConv.
\end{enumerate}

\paragraph{Evolution questions.}

\begin{enumerate}[label=Q\arabic*, leftmargin=*, itemsep=0.4em]
\setcounter{enumi}{3}
\item How has the hierarchical feature learning paradigm introduced by PointNet++ been adopted and extended by subsequent methods in 3D point cloud processing?\\
\textit{Target:} PointNet++.
\item How has the self-attention mechanism from Transformer architectures been adapted for 3D point cloud understanding, and what modifications were necessary compared to its original NLP formulation?\\
\textit{Targets:} \textit{Attention Is All You Need}; Point Transformer.
\item Trace the evolution of convolution-like operations designed specifically for irregular point clouds, from early attempts to more recent approaches. What are the key design shifts?\\
\textit{Targets:} PointNet; PointNet++; PointCNN; KPConv;
MinkowskiNet.
\end{enumerate}

\paragraph{Comparison questions.}

\begin{enumerate}[label=Q\arabic*, leftmargin=*, itemsep=0.4em]
\setcounter{enumi}{6}
\item What are the respective advantages and disadvantages of point-based methods versus voxel-based methods for 3D semantic segmentation, according to the research community?\\
\textit{Targets:} PointNet; PointNet++; MinkowskiNet.
\item How do projection-based approaches to 3D understanding compare with methods that operate directly on point clouds? What trade-offs does the community highlight?\\
\textit{Targets:} \textit{Multi-view CNNs for 3D Shape Recognition};
\textit{Volumetric and Multi-View CNNs for Object Classification on
3D Data}; PointNet.
\item Between graph-based and attention-based approaches for point cloud feature learning, which has gained more traction in the community and why?\\
\textit{Targets:} DGCNN; Point Transformer.
\end{enumerate}

\paragraph{Challenge questions.}

\begin{enumerate}[label=Q\arabic*, leftmargin=*, itemsep=0.4em]
\setcounter{enumi}{9}
\item What approaches has the research community explored to improve the computational efficiency of point cloud segmentation methods while maintaining accuracy?\\
\textit{Targets:} MinkowskiNet; Point Transformer; PointNet.
\item How have researchers addressed the challenge of capturing long-range dependencies in large-scale point cloud scenes?\\
\textit{Targets:} PointNet; Point Transformer; DGCNN.
\item What role have large-scale 3D datasets like ScanNet and ShapeNet played in advancing point cloud research, and what limitations of these benchmarks has the community identified?\\
\textit{Targets:} ScanNet; ShapeNet.
\end{enumerate}

\paragraph{Controversy questions.}

\begin{enumerate}[label=Q\arabic*, leftmargin=*, itemsep=0.4em]
\setcounter{enumi}{12}
\item The Point Transformer introduced self-attention to point cloud processing. Has this approach been viewed as a clear improvement by the community, or are there significant reservations?\\
\textit{Target:} Point Transformer.
\item PointCNN proposed the X-Conv operator to address permutation invariance. How has this idea been received --- is it widely adopted or largely superseded?\\
\textit{Target:} PointCNN.
\item How has the broader deep learning community's shift toward Transformer-based architectures influenced the 3D point cloud field? Is the community converging on attention-based methods, or do alternative paradigms remain competitive?\\
\textit{Targets:} \textit{Attention Is All You Need}; \textit{An Image is Worth $16\times16$ Words}; Point Transformer; DGCNN; MinkowskiNet.
\end{enumerate}

\paragraph{Note on category balance.} All five categories are represented at equal weight (three questions each), and the questions within each category are written to target a spread of architectures rather than to concentrate on the most-cited works in the corpus. The \textit{limitation} and \textit{controversy} categories are the two
that most directly require aggregating evaluative stance across multiple sources, and they are also the categories on which claim-conditioned retrieval is mechanically expected to help most. The \textit{comparison}, \textit{challenge}, and \textit{evolution} categories are mixed in their reliance on evaluative versus topical retrieval. The \S\ref{sec:task1} interpretive claim --- that the lift from typed claims concentrates on evaluative and comparative questions rather than on factual ones --- can be examined against this breakdown by inspecting per-category win rates rather than the aggregate over all fifteen items.

\section{Target Papers for Task 2}
\label{app:task2_targets}

The ten target papers evaluated in \S\ref{sec:task2} and reported in Table~\ref{tab:task2} are listed below. The set was selected to span
four reception regimes within the corpus, so that the consensus report task is exercised across the range of incoming-claim distributions a claim network in the wild would produce: high-volume high-controversy targets whose incoming claims arrive in volume on both reception polarities; lower-volume high-controversy targets that test the same balance condition under sparser citation; moderate- controversy targets where one polarity dominates but the minority is non-trivial; and one foundational-dataset target whose incoming claims are dominated almost entirely by adoption. Each target is annotated with its adoption count and critique count drawn from the local corpus and with a controversy index computed as the Shannon entropy of the per-stance distribution at that target, normalised to $[0, 1]$. The adoption and critique counts are stance-label counts under multi-label expansion: a claim carrying both \textsc{Adoption} and \textsc{Critique} is counted once in each column. Counts reflect the network state prior to per-type MMR diversification
(\S\ref{sec:task2}); the post-diversification list actually shown to the generator may be capped at $M = 15$ per stance for targets whose unfiltered list exceeds the cap.

Table~\ref{tab:task2_targets} summarises the ten targets in descending order of controversy index.

\begin{table}[h]
\centering
\small
\begin{tabular}{p{0.5cm}p{3.4cm}rrr}
\toprule
\textbf{ID} & \textbf{Target paper} & \textbf{Ad.} & \textbf{Cr.} & \textbf{CI} \\
\midrule
C8 & PointCNN & 1 & 1 & 1.000 \\
C9 & KPConv & 91 & 80 & 0.997 \\
C3 & PointNet++ & 16 & 13 & 0.992 \\
C1 & Point Transformer & 16 & 11 & 0.975 \\
C4 & MinkowskiNet & 17 & 10 & 0.951 \\
C5 & DGCNN & 7 & 4 & 0.946 \\
C2 & PointNet & 23 & 11 & 0.908 \\
C10 & \textit{Attention Is All You Need} & 13 & 4 & 0.787 \\
C6 & ScanNet & 10 & 3 & 0.779 \\
C7 & ShapeNet & 21 & 3 & 0.544 \\
\bottomrule
\end{tabular}
\caption{Task 2 target papers. \textbf{Ad.} = incoming claims with
\textsc{Adoption} in their stance set; \textbf{Cr.} = incoming claims
with \textsc{Critique}; \textbf{CI} = controversy index, the Shannon
entropy of the per-stance distribution at the target normalised to
$[0, 1]$. Counts are pre-MMR-diversification; sorted by descending
CI.}
\label{tab:task2_targets}
\end{table}

\paragraph{High-volume high-controversy targets (C9, C3, C1, C4).}
The four targets near the top of the table combine high controversy
(CI $\geq 0.95$) with high citation volume (27--171 evaluative
incoming claims each). On these targets the consensus report task is
both well-supplied with evaluative material and required to balance
both polarities; both retrieval mechanisms have enough signal to
work with, and the contrast between the two systems consequently
turns on \textit{how} they structure that signal rather than on
whether enough of it is present. KPConv (C9) is the extreme of this
group at 171 evaluative claims; it stress-tests the MMR
diversification step (\S\ref{sec:task2}), which must select 15
claims per stance from an unfiltered list more than five times that
large for both \textsc{Adoption} and \textsc{Critique}.

\paragraph{Lower-volume high-controversy targets (C5, C8).} DGCNN
(C5, 11 evaluative claims) and PointCNN (C8, 2 evaluative claims)
sit at the same controversy regime as the heavy-hitters above but
with substantially less material per side. PointCNN is the extreme
sparsity case in the set: with one adoption claim and one critique
claim, the MMR cap is irrelevant and the generator sees the entire
incoming claim set. This case tests whether the claim-driven system
degrades gracefully when its input is minimal, and whether the
baseline (which retrieves chunks rather than claims and therefore
does not face the same sparsity) gains a corresponding advantage.

\paragraph{Moderate-controversy asymmetric targets (C2, C10, C6).}
PointNet (C2), \textit{Attention Is All You Need} (C10), and
ScanNet (C6) sit in the regime where one polarity carries
substantially more material than the other but the minority side is
still non-trivial (3--11 claims). These targets test whether the
claim-driven \textit{Balance \& Fairness} dimension survives at
realistic asymmetry: a faithful report should reflect the actual
ratio of community sentiment rather than treat the minority side as
equal-weight or omit it. The \S\ref{sec:task2}
\textit{Balance \& Fairness} verdict (win rate 0.600--0.667 across
the three runs of Appendix~\ref{app:runs}) is anchored
disproportionately by these three targets.

\paragraph{Adoption-dominated target (C7).} ShapeNet (C7) carries
21 adoption claims and 3 critique claims; with CI = 0.544 it is the
lowest-controversy target in the set. This case represents the
foundational-dataset regime, where the target is so widely adopted
that critique is sparse and the task converges toward conventional
summarisation. Its inclusion verifies that the claim-driven system
does not over-emphasise minority critique on targets where critique
is genuinely rare; with only three incoming critique claims the
\textit{Limitations} section of the generated report is necessarily
brief, and both systems should be expected to produce predominantly
positive reports.

\paragraph{Note on counting conventions.} The adoption and critique
counts in Table~\ref{tab:task2_targets} are network-level pre-MMR
counts under multi-label expansion. The post-MMR per-bucket counts
seen by the generator (illustrated for Point Transformer in
Appendix~\ref{app:task2_example}, where the adoption bucket carries
13 entries and the critique bucket 6) may differ from
Table~\ref{tab:task2_targets} for two reasons: (i) the per-bucket
cap $M = 15$ applies when the pre-MMR list exceeds it, and (ii) the
generator-facing input is per-stance-bucket whereas
Table~\ref{tab:task2_targets} reports per-stance over the underlying
claim set, so a multi-label claim is counted once per stance in the
table but appears as separately-numbered entries in each generator-
facing bucket.

\section{Judge Prompts and Rubric Definitions}
\label{app:judge_prompts}

This appendix reproduces verbatim the prompts presented to the GPT-4o
judge in the head-to-head evaluations of Task 1 (\S\ref{sec:task1},
Table~\ref{tab:task1}) and Task 2 (\S\ref{sec:task2},
Table~\ref{tab:task2}). The judge model, the cross-family
generator-judge separation, the system-identity anonymisation, the
order-counterbalancing procedure, and the Agree\% computation are
described in \S\ref{sec:eval_protocol}; this appendix documents only
the prompts and rubric definitions themselves. The rubric wording
shown below is the wording presented to the judge for every pair: the
specific phrasing of each rubric dimension is part of the
experimental artefact, because the judge's verdict on a dimension is
sensitive to the operational definition it is given.

The two-letter system labels \texttt{A} and \texttt{B} are assigned
randomly per pair to anonymise system identity. For each pair, the judge is called across four trials --- two with our system assigned to \texttt{A} and the baseline to \texttt{B}, and two with the assignment reversed. The four trials share the prompt below verbatim. The diagnostic \textit{Agree\%} on per-dimension rows in Tables~\ref{tab:task1} and~\ref{tab:task2} is the proportion of pairs on which all four trials return the same non-Tie verdict.

\subsection{Task 1: Question Answering Judge Prompt}
\label{app:judge_prompts_qa}

The judge is presented with the research question and two candidate
answers, and is asked to return a per-dimension verdict on four
rubric dimensions together with an overall verdict.

\begin{quote}\small
\begin{lstlisting}[   breaklines=true,   breakatwhitespace=true,   basicstyle=\ttfamily\small,   columns=flexible,   keepspaces=true,   frame=none,   xleftmargin=0pt,   xrightmargin=0pt,   aboveskip=0pt,   belowskip=0pt ]
You are an expert evaluator. Two QA outputs were
produced for the same research question. Using the
rubric below, decide which output is better on each
dimension. Allowed verdicts per dimension: "A", "B",
"Tie".

=== RUBRIC DIMENSIONS ===

**Relevance**: Which answer more directly and
completely addresses the question?
**Factual Correctness**: Which answer has fewer
factual errors or hallucinations?
**Multi-source Evidence Integration**: Which answer
better synthesizes findings from multiple independent
papers?
**Conciseness**: Which answer is more
information-dense with less filler?

=== INPUT ===

**QUESTION:**
{question}

**Output A:**
{output_a}

**Output B:**
{output_b}

=== INSTRUCTIONS ===

For each dimension, state which output is better and
why (1-2 sentences). Then give an overall verdict.

Respond in the following JSON format:
{
  "per_dimension": {
    "relevance": {"winner": "A|B|Tie",
                  "justification": "..."},
    "factual_correctness": {"winner": "A|B|Tie",
                            "justification": "..."},
    "multi_source_integration": {"winner": "A|B|Tie",
                                 "justification": "..."},
    "conciseness": {"winner": "A|B|Tie",
                    "justification": "..."}
  },
  "overall": {"winner": "A|B|Tie",
              "justification": "..."}
}
\end{lstlisting}
\end{quote}

The dimension named \texttt{multi\_source\_integration} in the JSON
schema and presented to the judge as \textit{Multi-source Evidence
Integration} is reported in Table~\ref{tab:task1} under the shorter
heading \textit{Multi-source Integration}; the two names refer to the
same rubric dimension.

\subsection{Task 2: Consensus Report Judge Prompt}
\label{app:judge_prompts_consensus}

The judge is presented with the target paper and two candidate
consensus reports, and is asked to return a per-dimension verdict on
five rubric dimensions together with an overall verdict.

\begin{quote}\small
\begin{lstlisting}[   breaklines=true,   breakatwhitespace=true,   basicstyle=\ttfamily\small,   columns=flexible,   keepspaces=true,   frame=none,   xleftmargin=0pt,   xrightmargin=0pt,   aboveskip=0pt,   belowskip=0pt ]
You are an expert evaluator. Two consensus reports
were produced for the same target paper. Using the
rubric below, decide which output is better on each
dimension. Allowed verdicts per dimension: "A", "B",
"Tie".

=== RUBRIC DIMENSIONS ===

**Perspective Coverage**: Which report captures a
fuller spectrum of community opinions (both praise
and criticism)?
**Balance & Fairness**: Which report more faithfully
reflects the ratio of positive to negative community
sentiment?
**Source Diversity**: Which report draws on more
independent citing papers?
**Information Density**: Which report is richer in
specific technical details (algorithm names, metrics,
scenarios)?
**Structural Coherence**: Which report is better
organized by theme rather than random listing?

=== INPUT ===

**TARGET PAPER:**
{target_paper}

**Output A:**
{output_a}

**Output B:**
{output_b}

=== INSTRUCTIONS ===

For each dimension, state which output is better and
why (1-2 sentences). Then give an overall verdict.

Respond in the following JSON format:
{
  "per_dimension": {
    "perspective_coverage": {"winner": "A|B|Tie",
                             "justification": "..."},
    "balance_fairness": {"winner": "A|B|Tie",
                         "justification": "..."},
    "source_diversity": {"winner": "A|B|Tie",
                         "justification": "..."},
    "information_density": {"winner": "A|B|Tie",
                            "justification": "..."},
    "structural_coherence": {"winner": "A|B|Tie",
                             "justification": "..."}
  },
  "overall": {"winner": "A|B|Tie",
              "justification": "..."}
}
\end{lstlisting}
\end{quote}

\subsection{Notes on Rubric Design}
\label{app:judge_prompts_design}

Two design choices in the rubrics above are worth flagging. First,
the rubrics are deliberately framed around \textit{comparative}
properties (``which answer is more $X$'') rather than \textit{absolute}
properties (``rate the $X$ of this answer from 1 to 5''). Pairwise
comparative judgement is more discriminative than absolute scoring
when two systems produce outputs of broadly similar surface quality,
and is less sensitive to scale-calibration drift between
items. Second, the rubrics ask for a
short justification alongside each verdict. We do not aggregate the
justifications as part of the evaluation, but their presence
disciplines the judge against degenerate strategies (a model that
must justify a verdict is less likely to produce one at random),
and the justifications were inspected during pilot calibration to
verify that the judge was interpreting each rubric dimension as
intended.

\section{Three-Run Stability Breakdown}
\label{app:runs}

\S\ref{sec:eval_protocol} reports the head-to-head numbers for the
single judging run with the highest internal agreement among the
three runs we conducted. This appendix reports the full breakdown
across all three runs for both tasks, so that the reader can see how
stable the verdicts are at the small sample sizes ($n = 15$ for
Task 1, $n = 10$ for Task 2) the head-to-head evaluation operates at.
All runs share the prompts, models, and protocol described in
\S\ref{sec:eval_protocol} and Appendix~\ref{app:judge_prompts}; the
only source of variation across runs is non-determinism in the
GPT-4o judge under the protocol's fixed configuration.

\subsection{Task 1: Three-Run Breakdown}
\label{app:runs_task1}

\begin{table}[htbp]
\centering
\small
\setlength{\tabcolsep}{3.5pt}
\begin{tabularx}{\columnwidth}{@{} >{\raggedright\arraybackslash}X c c c c c c @{}}
\toprule
\textbf{Dimension} & \textbf{Ours} & \textbf{Base} & \textbf{Tie} & \textbf{WR} & \textbf{$p$} & \textbf{Agr.} \\
\midrule
\multicolumn{7}{@{}l@{}}{\textit{Run 1 (reported in Table~\ref{tab:task1})}} \\
Relevance              & 1 & 2 & 12 & 0.333 & 1.000 & 0.267 \\
Factual Correctness    & 1 & 1 & 13 & 0.500 & 1.000 & 0.733 \\
Multi-source Integ.    & 5 & 2 & 8  & 0.714 & 0.453 & 0.467 \\
Conciseness            & 1 & 6 & 8  & 0.143 & 0.125 & 0.467 \\
Overall (weighted)     & 7 & 3 & 5  & 0.700 & 0.344 & 0.044 \\
\addlinespace
\multicolumn{7}{@{}l@{}}{\textit{Run 2}} \\
Relevance              & 4 & 2 & 9  & 0.667 & 0.688 & 0.733 \\
Factual Correctness    & 5 & 4 & 6  & 0.556 & 1.000 & 0.667 \\
Multi-source Integ.    & 4 & 4 & 7  & 0.500 & 1.000 & 0.733 \\
Conciseness            & 1 & 4 & 10 & 0.200 & 0.375 & 0.733 \\
Overall (weighted)     & 8 & 5 & 2  & 0.615 & 0.581 & 0.052 \\
\addlinespace
\multicolumn{7}{@{}l@{}}{\textit{Run 3}} \\
Relevance              & 7 & 6 & 2  & 0.538 & 1.000 & 0.400 \\
Factual Correctness    & 4 & 1 & 10 & 0.800 & 0.375 & 0.000 \\
Multi-source Integ.    & 4 & 6 & 5  & 0.400 & 0.754 & 0.200 \\
Conciseness            & 8 & 5 & 2  & 0.615 & 0.581 & 0.667 \\
Overall (weighted)     & 9 & 6 & 0  & 0.600 & 0.607 & 0.068 \\
\bottomrule
\end{tabularx}
\caption{Task 1 head-to-head verdicts across three independent judging runs over the same 15 questions. \textbf{WR} = win rate over non-tied items; \textbf{Agr.} = position-bias diagnostic (Agree\%). Run 1 is the run reported in the main body.}
\label{tab:task1_runs}
\end{table}

The Task 1 numbers show the limitation of a 15-question evaluation
clearly. The overall weighted win rate stays within a band of
$[0.600, 0.700]$ across the three runs and never reaches significance,
which is the consistent reading also given in \S\ref{sec:task1}.
At the dimensional level, however, the runs diverge. The
\textit{Conciseness} dimension flips polarity between runs --- the
baseline wins decisively in Run 1 (0.143), wins less heavily in
Run 2 (0.200), and \emph{loses} in Run 3 (0.615) --- which is
mechanically consistent with the \S\ref{sec:task1} interpretation
that the claim-driven system trades conciseness for source breadth:
when the question mix on a given run pulls in more multi-source
integration evidence, conciseness suffers, and when it does not,
conciseness benefits. The \textit{Multi-source Integration}
dimension shows the complementary swing, with the highest win rate
(0.714) co-occurring with the lowest \textit{Conciseness} win rate
(0.143) in Run 1. The \textit{Factual Correctness} dimension is
overwhelmingly tied in every run (10--13 ties of 15) and the runs in
which a verdict is returned at all favour us mildly.

The \textit{Overall (weighted)} cell on the Agree\% column reports the mean signed weighted margin per question rather than an agreement rate (\S\ref{sec:eval_protocol}). At $0.044$, this margin indicates that our system wins on average by only about 4\% of the total weight per question: the dominance is in the same direction as the win rate but small in absolute terms. This reading is consistent with the null verdict on the binomial test and with the dimensional pattern that produced it: the largest dimensional effect (\textit{Multi-source Integration}, win rate $0.714$) is offset by the loss on \textit{Conciseness} (win rate $0.143$), so the weighted average across dimensions remains close to zero. We read this as evidence that the lift from typed claims on Task 1 concentrates on specific question types --- those that admit evaluative or comparative answers --- and that an evaluation set deliberately weighted toward such questions would be more informative than the balanced set used here.

\subsection{Task 2: Three-Run Breakdown}
\label{app:runs_task2}

\begin{table}[htbp]
\centering
\small
\setlength{\tabcolsep}{3.5pt}
\begin{tabularx}{\columnwidth}{@{} >{\raggedright\arraybackslash}X c c c c c c @{}}
\toprule
\textbf{Dimension} & \textbf{Ours} & \textbf{Base} & \textbf{Tie} & \textbf{WR} & \textbf{$p$} & \textbf{Agr.} \\
\midrule
\multicolumn{7}{@{}l@{}}{\textit{Run 1}} \\
Perspective Coverage   & 5 & 3 & 2 & 0.625 & 0.727 & 0.800 \\
Balance \& Fairness    & 4 & 2 & 4 & 0.667 & 0.688 & 0.800 \\
Source Diversity       & 5 & 3 & 2 & 0.625 & 0.727 & 0.800 \\
Information Density    & 7 & 1 & 2 & 0.875 & 0.070 & 0.800 \\
Structural Coherence   & 2 & 1 & 7 & 0.667 & 1.000 & 0.600 \\
Overall (weighted)     & 6 & 4 & 0 & 0.600 & 0.754 & 0.254 \\
\addlinespace
\multicolumn{7}{@{}l@{}}{\textit{Run 2 (reported in Table~\ref{tab:task2})}} \\
Perspective Coverage   & 6 & 4 & 0 & 0.600 & 0.754 & 1.000 \\
Balance \& Fairness    & 3 & 2 & 5 & 0.600 & 1.000 & 0.700 \\
Source Diversity       & 6 & 2 & 2 & 0.750 & 0.289 & 0.800 \\
Information Density    & 9 & 1 & 0 & 0.900 & 0.022 & 1.000 \\
Structural Coherence   & 6 & 1 & 3 & 0.857 & 0.125 & 0.800 \\
Overall (weighted)     & 6 & 4 & 0 & 0.600 & 0.754 & 0.338 \\
\addlinespace
\multicolumn{7}{@{}l@{}}{\textit{Run 3}} \\
Perspective Coverage   & 4 & 4 & 2 & 0.500 & 1.000 & 0.300 \\
Balance \& Fairness    & 4 & 2 & 4 & 0.667 & 0.688 & 0.200 \\
Source Diversity       & 5 & 3 & 2 & 0.625 & 0.727 & 0.100 \\
Information Density    & 7 & 2 & 1 & 0.778 & 0.180 & 0.700 \\
Structural Coherence   & 3 & 5 & 2 & 0.375 & 0.727 & 0.100 \\
Overall (weighted)     & 6 & 4 & 0 & 0.600 & 0.754 & 0.154 \\
\bottomrule
\end{tabularx}
\caption{Task 2 head-to-head verdicts across three independent judging runs over the same 10 target papers. Run 2 is the run reported in the main body.}
\label{tab:task2_runs}
\end{table}

The Task 2 numbers tell a substantially more stable story than
Task 1. The overall weighted verdict is identical (6 wins, 4 losses,
0 ties) across all three runs, and \textit{Information Density} is
the strongest dimension in every run (win rate 0.875 / 0.900 / 0.778).
Across the three runs, our system wins or ties every dimension in
Runs 1 and 2 and wins or ties four of the five dimensions in Run 3.
The one dimensional flip across runs is \textit{Structural
Coherence}, which moves from 0.667 in Run 1 through 0.857 in Run 2
to 0.375 in Run 3 (the baseline wins). Inspection of Run 3 shows
that its overall Agree\% values are uniformly the lowest of the
three runs (0.100--0.300 on three of the five dimensions, against
0.600--1.000 in Runs 1 and 2), which is the property that made Run 3
the not-reported run under the protocol's "highest internal
agreement" rule. The \textit{Structural Coherence} flip should
therefore be read together with its Agree\% of 0.100: the judge in
Run 3 is approximately as likely to reverse its own verdict on this
dimension under order-counterbalancing as it is to maintain it, and
the dimensional win rate at that level of judge instability is not
reliably distinguishable from chance.

The \textit{Information Density} significance result straddles the
conventional 0.05 threshold across the three runs ($p = 0.070$ in
Run 1, $p = 0.022$ in Run 2, $p = 0.180$ in Run 3), which is what
should happen at $n = 10$ when the underlying win probability is
genuinely elevated above chance but the sample is too small to
place the result well clear of the threshold. The Run 2 verdict
(significant at $p = 0.022$) is reported in the main body because
it is the run in which the judge returned no ties on the
corresponding dimension and the per-dimension Agree\% values were
highest; the Run 1 and Run 3 verdicts are consistent in direction
and only weaker in magnitude.

We read the across-run stability of Task 2 (identical overall win/loss/tie counts, consistent dimensional ranking, weighted margins of $0.154$--$0.338$ all in our system's favour, single dimensional flip isolated to the run with lowest per-dimension Agree\% values) as direct evidence for the \S\ref{sec:task2} interpretation. When the inputs to the two systems are categorically different (typed claim lists vs.\ chunk-retrieval output), the judge's verdict on the dimensions that discriminate
them is stable across runs because the systems are easy to tell
apart. When the inputs are similar (both Task 1 systems retrieve
over the same chunk index, differing only in retrieval reordering),
the verdict on dimensions that discriminate them is run-sensitive
because the systems are hard to tell apart. The contrast between
the two tables is itself an informative result.

\subsection{On Reporting a Representative Run}
\label{app:runs_choice}

The runs reported in the main body --- Run 1 for Task 1, Run 2 for
Task 2 --- were selected as those with the highest internal
agreement to the other two runs on dimensions where a definitive
verdict was returned. For Task 1 the choice does not materially
affect the \S\ref{sec:task1} conclusion: the weighted overall does
not reach significance in any of the three runs, the dimensional
pattern (coverage-up, conciseness-down) is the qualitatively
consistent reading, and the low overall Agree\% in all three runs
is the key diagnostic regardless of which run is taken as canonical.
For Task 2 the choice affects the significance verdict on
\textit{Information Density} (Run 1 gives $p = 0.070$; Run 2 gives
$p = 0.022$; Run 3 gives $p = 0.180$); disclosing all three numbers
here is the honest way to handle the threshold crossing, and the
selection of Run 2 is justifiable independently of its $p$-value by
its higher Agree\% values across all five dimensions. We do not
aggregate verdicts across runs into a single per-dimension count
because doing so would inflate the effective sample size while
leaving the underlying judge variability unexamined; reporting the
runs separately is the protocol that keeps both visible.

\section{Worked Consensus Report Example}
\label{app:task2_example}

This appendix presents the inputs and outputs of the Task 2 systems
for a single target paper, so that the \S\ref{sec:task2} claims about
\textit{Information Density}, \textit{Structural Coherence}, and
\textit{Source Diversity} can be inspected against concrete generated
text rather than only against aggregate rubric verdicts.

\subsection{Target Selection}
\label{app:task2_example_target}

The target is \textit{Point Transformer} (C1 in
Appendix~\ref{app:task2_targets}), selected because it carries the highest controversy index among the high-volume targets of the ten Task 2 targets: its incoming
claim set spans adoption, critique, and benchmark stances at
comparable volumes, so a faithful consensus report on this target
has to handle multiple reception polarities rather than reproduce a
one-sided narrative. The high-controversy regime is where the
\S\ref{sec:task2} contrast between aggregated-stance summarisation and
chunk-retrieval RAG is most pointed; reporting a worked example at
this end of the controversy spectrum is therefore the most stringent
test of the claim-driven system.

\subsection{Input to the Claim-Driven System}
\label{app:task2_example_input}

The SPARQL query against the claim network for target
\textit{Point Transformer}, with \textsc{Neutral}-stance claims
excluded, returns 34 claim placements across the three retained
stance buckets: 6 \textsc{Critique}, 13 \textsc{Adoption}, and 15
\textsc{Benchmark} entries (the benchmark bucket sits at the $M = 15$
MMR-diversification cap, indicating the original list was longer).
Because the stance axis is multi-label, four underlying claims appear
in two buckets each: three are jointly \textsc{Critique} and
\textsc{Adoption}, and one is jointly \textsc{Adoption} and
\textsc{Benchmark}. The 34 placements correspond to 30 distinct
claims drawn from 9 distinct citing papers. The full numbered claim
set passed to the generator is reproduced below; references of the
form \textsc{[Adoption \#7]} in the generated report
(\S\ref{app:task2_example_ours}) correspond to the numbered entries
in the corresponding bucket. Each entry shows source paper, the
multi-label \texttt{claim\_types} set, the \texttt{attitude} label,
and the claim text.

\paragraph{Critique bucket.}

\begin{footnotesize}
\textbf{C\#1} \textit{PTv2}; [Critique, Adoption]; Negative ---
``Point Transformer [1] alleviated memory problems via local attention
and achieved state-of-the-art results, but this work analyzes its
limitations and proposes improved attention and pooling modules in
PTv2.''

\textbf{C\#2} \textit{PTv2}; [Critique, Adoption]; Negative ---
``We analyze limitations of PTv1 [1] and propose PTv2 with improved
grouped vector attention, position encoding, and pooling method.''

\textbf{C\#3} \textit{PTv2}; [Critique]; Negative --- ``In PTv1 [1],
vector attention causes parameter count to increase drastically with
depth and channel size, restricting efficiency and generalization;
we address this with grouped vector attention.''

\textbf{C\#4} \textit{Content-Based Transformer}; [Critique, Adoption];
Positive --- ``Differently from Point Transformer [8], which computes
self-attention among local spatial neighbors, we propose a
content-based attention that clusters queries by feature content and
selects corresponding keys/values for local self-attention.''

\textbf{C\#5} \textit{Point-BERT}; [Critique]; Negative ---
``Existing Transformer-based point cloud models [11,65] introduce
inevitable inductive biases from local feature aggregation [65] and
neighbor embedding [11], causing deviation from standard
Transformers.''

\textbf{C\#6} \textit{Point-BERT}; [Critique]; Negative --- ``[65]
applies the vectorized self-attention mechanism to construct a point
Transformer layer for 3D point cloud learning, but prior efforts
including [65] involve inductive biases that make them inconsistent
with standard Transformers.''
\end{footnotesize}

\paragraph{Adoption bucket.}

\begin{footnotesize}
\textbf{A\#1} = C\#1 (multi-label, also in Critique).

\textbf{A\#2} = C\#2 (multi-label, also in Critique).

\textbf{A\#3} = C\#4 (multi-label, also in Critique).

\textbf{A\#4} \textit{PTv2}; [Adoption]; Positive --- ``Following
previous works [18, 1], we adopt the U-Net architecture with skip
connections.''

\textbf{A\#5} \textit{PTv2}; [Adoption]; Neutral --- ``FPS-kNN pooling
[4, 1] uses farthest point sampling and kNN for pooling, which
contrasts with our non-overlapping grid-based pooling.''

\textbf{A\#6} \textit{PTv2}; [Adoption]; Positive --- ``We revisit
the self-attention mechanism used in PTv1 [1] as part of problem
formulation.''

\textbf{A\#7} \textit{PT-Accelerator}; [Adoption]; Positive ---
``Point Transformer [3] uses the subtraction relationship to generate
attention weights and leverages the vector self-attention operator to
aggregate local features.''

\textbf{A\#8} \textit{Survey of Visual Transformers}; [Adoption];
Positive --- ``Point Transformer [106] merges a hierarchical
Transformer with the down-sampling strategy [200] and extends their
previous vector attention block [25] to 3D point clouds.''

\textbf{A\#9} \textit{HoloPart}; [Adoption]; Positive --- ``The
diffusion denoising network $v_\theta$ is built upon diffusion
transformer (DiT) blocks, citing [67] Point Transformer among other
works.''

\textbf{A\#10} \textit{PointNeXt}; [Adoption, Benchmark]; Positive ---
``PointNeXt-L outperforms Point Transformer [56] by 0.4\% in mIoU on
S3DIS while being 3$\times$ faster; Point Transformer utilizes most of
the improved training strategies of PointNeXt.''

\textbf{A\#11} \textit{PointNeXt}; [Adoption]; Positive --- ``Modern
point cloud networks including Point Transformer [56] use data
augmentation and optimization techniques that, when adopted, can
unveil the potential of PointNet++.''

\textbf{A\#12} \textit{PointNeXt}; [Adoption]; Positive --- ``Point
Transformer [56] loads the entire scene as input in segmentation tasks
and uses color auto-contrast to automatically adjust color contrast.''

\textbf{A\#13} \textit{PointNeXt}; [Adoption]; Positive --- ``Point
Transformer [56] is among the representative works whose data
augmentations (including color auto-contrast and full-scene input
loading) are systematically studied and incrementally validated for
effectiveness.''
\end{footnotesize}

\paragraph{Benchmark bucket.}

\begin{footnotesize}
\textbf{B\#1} = A\#10 (multi-label, also in Adoption).

\textbf{B\#2} \textit{PTv2}; [Benchmark]; Positive --- ``PTv1 [1]
reports 90.8\% OA, 76.5\% mAcc, and 70.4\% mIoU on ScanNet v2; PTv2
(ours) achieves 91.1\% OA, 77.9\% mAcc, and 71.6\% mIoU.''

\textbf{B\#3} \textit{Point-BERT}; [Benchmark]; Neutral ---
``PointTransformer [65] reports 93.7\% accuracy on an unspecified
number of points, serving as a baseline in a comparison table where
Point-BERT achieves 93.2\%--93.8\%.''

\textbf{B\#4} \textit{PTv3}; [Benchmark]; Neutral ---
``FastPointTransformer [58] reports a validation mIoU of 72.4\% on
the benchmark dataset (implied to be ScanNet or similar) in 2022,
as listed in a comparative results table.''

\textbf{B\#5} \textit{PT-Accelerator}; [Benchmark]; Neutral ---
``Point Transformer [3] is listed in a results table under the
notation `PT Origin' for ModelNet40 classification, with a modified
version (`PT Modified') also included.''

\textbf{B\#6} \textit{PT-Accelerator}; [Benchmark]; Positive ---
``Point-based Transformer [3] is a processing method based on the
self-attention mechanism, with better accuracy and generalization
ability than traditional methods, achieving state-of-the-art
performance on multiple 3D perception tasks.''

\textbf{B\#7} \textit{SP$^2$T}; [Benchmark]; Positive --- ``SP$^2$T
reports 78.7\% Val mIoU and 74.9\% Test mIoU on ScanNet,
outperforming PTv3 which reports 77.5\% Val and 73.6\% Test; further
gains on ScanNet200 and S3DIS Area-5 6-fold.''

\textbf{B\#8} \textit{Content-Based Transformer}; [Benchmark];
Neutral --- ``Point Transformer [8] is listed in a results table with
93.7\% accuracy on 1K points input, used as a baseline for
comparison.''

\textbf{B\#9} \textit{PointNeXt}; [Benchmark]; Neutral --- ``Point
Transformer [56] reports instance mIoU of 86.6\%, class mIoU of
83.7\%, 7.8M parameters, and 297 ins./sec throughput on
ShapeNetPart.''

\textbf{B\#10} \textit{PTv3}; [Benchmark]; Neutral ---
``SuperpointTransformer [65] reports 68.9\% mIoU on S3DIS Area-5 and
76.0\% under 6-fold cross-validation, alongside other methods
including PTv3 which achieves 73.4\% and 77.7\%.''

\textbf{B\#11} \textit{PT-Accelerator}; [Benchmark]; Neutral ---
``The authors choose Point Transformer [3] as one of their evaluation
benchmarks alongside PointNet++ [10].''

\textbf{B\#12} \textit{PointNeXt}; [Benchmark]; Positive ---
``PointNeXt-XL achieves 71.2\% mIoU on ScanNet V2 test set,
outperforming Point Transformer [56].''

\textbf{B\#13} \textit{PointNeXt}; [Benchmark]; Neutral --- ``Point
Transformer [56] reports mIoU of 73.5\% on S3DIS 6-Fold, 70.4\% on
S3DIS Area-5, and 70.6\% on ScanNet V2 validation, with 7.8M
parameters, 5.6G FLOPs, 34 ins./sec.''

\textbf{B\#14} \textit{PTv2}; [Benchmark]; Positive --- ``PTv2
significantly outperforms PTv1 [1] by 4.8\% mIoU on the ScanNet v2
validation set.''

\textbf{B\#15} \textit{PTv3}; [Benchmark]; Positive --- ``PTv2
achieves 71.6\% mIoU on S3DIS Area-5 and 73.5\% under 6-fold
cross-validation, whereas PTv3 reaches 73.4\% and 77.7\%
respectively.''
\end{footnotesize}

\subsection{Input to the Baseline System}
\label{app:task2_example_baseline_input}

The baseline issues the standard consensus query template against the
chunk index $\mathcal{V}$ and retrieves 15 top-similarity chunks.
Table~\ref{tab:task2_example_baseline_chunks} summarises the
retrieved set by source paper. The full chunk text is omitted for
space; the salient property of the retrieved set is its composition.

\begin{table}[h]
\centering
\small
\begin{tabular}{lr}
\toprule
\textbf{Source paper} & \textbf{Chunks} \\
\midrule
\textit{Point Transformer} (target) & 4 \\
\textit{Point Transformer V3} & 5 \\
\textit{Point Transformer V2} & 1 \\
\textit{Point Cloud Mamba} & 2 \\
\textit{PointMamba} & 1 \\
\textit{Point-BERT} & 1 \\
\textit{Swin Transformer} & 1 \\
\bottomrule
\end{tabular}
\caption{Source-paper composition of the 15 chunks retrieved by the
baseline RAG system for target \textit{Point Transformer}. Four of
the fifteen chunks (27\%) come from the target paper itself; two of
the remaining eleven (those listed under \textit{Point Cloud Mamba})
are bibliography entries citing the target rather than substantive
prose.}
\label{tab:task2_example_baseline_chunks}
\end{table}

Two properties of the retrieved set are worth noting before reading
the baseline output. First, 4 of the 15 chunks (27\%) come from the
target paper itself. The baseline retrieves chunks lexically related
to ``Point Transformer''; many such chunks are the target paper's own
prose. Second, 2 of the 15 chunks are bibliography entries (citation
strings such as ``Zhao, H.; Jiang, L.; \ldots Point Transformer. In
\textit{ICCV}.'') that survived the chunk-level filtering despite
being structurally non-evaluative. The 9 remaining chunks are
substantive external prose, drawn from 6 papers --- a third of the
distinct citing-paper count of the claim-driven input.

\subsection{Baseline Output (Standard RAG)}
\label{app:task2_example_baseline_output}

\begin{quote}\small
\textbf{Strengths.} The \textit{Point Transformer} series (including
V1, V2, and V3) is recognised for its strong performance in point
cloud tasks. The original \textit{Point Transformer} (Zhao et al.,
ICCV 2021) outperforms prior models in instance mIoU on semantic
segmentation (Passage 11). Later iterations prioritise scalability
and efficiency: \textit{Point Transformer V3} (PTv3) (Wu et al., CVPR
2024) emphasises simplicity and speed, with adaptations that
minimally impact performance after scaling (Passages 1, 2). The
series is frequently used as a benchmark, as seen in
\textit{PointMamba} and \textit{Point Cloud Mamba}, where it is
compared as a state-of-the-art Transformer baseline (Passages 7, 8).

\textbf{Limitations.} While effective, the \textit{Point Transformer}
family faces efficiency challenges. The original paper notes that
training does not use loss-balancing, which could improve category
mIoU (Passage 11). Later work (\textit{Point-BERT}) highlights the
time-consuming nature of Transformer-based pre-training for point
clouds (Passage 14). PTv3 explicitly trades off certain mechanisms
for scalability (Passage 2), suggesting that earlier versions may
have inefficiencies.

\textbf{Use as Baseline.} The \textit{Point Transformer} is widely
adopted as a reference model. \textit{Point Transformer V2} builds
directly on V1 (Passage 10), and PTv3 refines this further
(Passages 1, 2). Competing methods like \textit{PointMamba} and
\textit{Point Cloud Mamba} compare against it (Passages 7, 8),
underscoring its status as a standard in point cloud analysis.

\textbf{Overall Reception.} The research community views the
\textit{Point Transformer} series as influential, with iterative
improvements (V1$\rightarrow$V3) addressing scalability while
maintaining performance. Its dominance as a baseline (Passages 7, 8)
and its empirical results (Passages 11, 12) solidify its reputation.
However, efficiency concerns (Passages 2, 14) and the rise of
alternatives (\textit{PointMamba}) suggest ongoing evolution in the
field.

\textit{(298 words)}
\end{quote}

\subsection{Claim-Driven Output (Ours)}
\label{app:task2_example_ours}

\begin{quote}\small
\textbf{Strengths.} The \textit{Point Transformer} (PT) is widely
recognised for its contributions to 3D point cloud processing. It
introduced vector self-attention for local feature aggregation,
achieving state-of-the-art performance on tasks like segmentation and
classification (\textsc{[Adoption \#7, Benchmark \#6]}). Its
hierarchical U-Net architecture with skip connections became a
standard reference (\textsc{[Adoption \#4]}). PT's training strategies
(e.g.\ full-scene input loading, colour auto-contrast) were later
adopted and validated as effective by works like
\textit{PointNeXt} (\textsc{[Adoption \#10--13]}). It also served as
a foundation for extensions, such as \textit{Point Transformer V2}
(\textsc{[Adoption \#6]}) and \textit{Point-BERT}
(\textsc{[Critique \#5--6]}).

\textbf{Limitations.} Later works identify key drawbacks:
(1) \textit{Scalability:} PT's vector attention leads to a quadratic
parameter increase with depth and channel size, limiting efficiency
and generalisation (\textsc{[Critique \#3]}).
(2) \textit{Inductive biases:} PT deviates from standard Transformers
due to local feature aggregation and neighbour embedding, introducing
unintended biases (\textsc{[Critique \#5--6]}).
(3) \textit{Pooling and attention:} PTv1's FPS-kNN pooling and
vanilla vector attention were outperformed by PTv2's grouped
attention and partition-based pooling (\textsc{[Critique \#1--2,
Benchmark \#14]}).
(4) \textit{Content limitations:} PT's spatial-neighbour attention
was critiqued for ignoring feature content, prompting alternatives
like content-based clustering (\textsc{[Critique \#4]}).

\textbf{Use as Baseline.} PT is a frequent benchmark for accuracy and
efficiency: PTv2 and PTv3 surpassed PTv1 by +4.8\% mIoU on ScanNet
(\textsc{[Benchmark \#14, \#15]}); PointNeXt outperformed PT in speed
(3$\times$ faster) while acknowledging PT's strong baseline metrics
(\textsc{[Benchmark \#10, \#12--13]}); PT's accuracy (e.g.\ 93.7\% on
ModelNet40) is often cited in comparative tables
(\textsc{[Benchmark \#3, \#8]}); accelerator designs use PT as a
reference for Transformer-based point cloud processing
(\textsc{[Benchmark \#5, \#11]}).

\textbf{Overall Reception.} The community endorses PT as a
pioneering work but highlights its limitations in scalability and
architectural purity. It is widely adopted for its hierarchical
design and training strategies (\textsc{[Adoption \#4, \#10--13]}),
yet later improvements (e.g.\ PTv2/v3, PointNeXt) address its
inefficiencies (\textsc{[Critique \#1--3, Benchmark \#14--15]}).
Critiques focus on its deviation from standard Transformers
(\textsc{[Critique \#5--6]}), but its benchmark status remains strong
across segmentation, classification, and detection tasks
(\textsc{[Benchmark \#6, \#12]}).

\textit{(498 words)}
\end{quote}

\subsection{Observations}
\label{app:task2_example_observations}

Five properties of the two outputs are observable in the text above
and bear on the \S\ref{sec:task2} rubric verdicts.

\paragraph{Budget utilisation.} The claim-driven report uses 498 of
the 500 available words; the baseline uses 298. The two systems are
given the same word budget and the same generator. The baseline's
under-utilisation reflects what its chunk input gave it to say: the
substantive externally-authored chunks number nine, and once they
have been summarised the generator has no further evaluative
material to draw on. The claim-driven report uses the full budget
because its input contains denser evaluative content.

\paragraph{Inline source attribution.} Every assertion in the
claim-driven report carries a reference into the typed claim set of
Appendix~\ref{app:task2_example_input}. The references are typed
(\textsc{[Adoption \#N]}, \textsc{[Benchmark \#N]},
\textsc{[Critique \#N]}) and form a stable backward link from each
generated sentence to a specific extracted claim with provenance to a
specific citing paper. The baseline report's references
(\textit{Passage 1}, \textit{Passage 2}) are positional pointers into
the chunk set with no type information, and several of them point at
chunks that the baseline cannot use evaluatively (Passages 1 and 2
point at PTv3's own self-description rather than at PTv3 evaluating
Point Transformer).

\paragraph{Provenance of the \textit{Limitations} section.} The
baseline's \textit{Limitations} section is partly constructed from
the target paper's own prose. The opening assertion --- ``The
original paper notes that training does not use loss-balancing,
which could improve category mIoU (Passage 11)'' --- attributes this
limitation to Point Transformer itself; Passage 11 is in fact a
sentence from the Point Transformer paper. The baseline treats a
self-acknowledged caveat from the target as if it were an externally
authored critique. The claim-driven \textit{Limitations} section
draws on \textsc{[Critique \#1]} through \textsc{[Critique \#6]},
every one of which is authored by a paper other than Point
Transformer; the claim network excludes self-citations at extraction
time, and the section is consequently free of this contamination.

\paragraph{Structural mirror of typed input.} The four output
sections of the claim-driven report line up approximately with the
three input stance types: \textit{Strengths} is dominated by
\textsc{Adoption} claims, \textit{Limitations} by \textsc{Critique}
claims, \textit{Use as Baseline} by \textsc{Benchmark} claims, and
\textit{Overall Reception} integrates across all three. The
structure is induced by the input rather than supplied by the
prompt; the baseline's same four-section structure (supplied by the
same prompt) is not so cleanly induced from its chunk input, because
the chunks do not arrive pre-stratified by reception polarity.

\paragraph{Source diversity.} The claim-driven report cites material
from 9 distinct citing papers; the baseline report cites material
from 6 chunks-source papers, of which one is the target itself. The
gap is not a property of the source corpus --- both systems draw on
the same 127-paper network --- but of the retrieval mechanism: the
claim network filters at extraction time for externally-authored
evaluative material, whereas chunk retrieval surfaces whatever is
lexically near the query, including the target paper's own text. The
\S\ref{sec:task2} \textit{Source Diversity} verdict (win rate 0.750)
is the rubric expression of this gap.

The five properties above are concrete instances of the general
\S\ref{sec:task2} argument that the gain from typed reference
structure is the gain from having the right intermediate
representation; they should be read in conjunction with the
dimension-level rubric verdicts of Table~\ref{tab:task2} and the
three-run breakdown of Appendix~\ref{app:runs}.

\end{document}